%% file: main.tex
\documentclass[12pt]{iopart}
\usepackage{graphicx}
\usepackage[utf8]{inputenc}
\usepackage[T1]{fontenc}
\usepackage{subfig}
\usepackage{graphicx}
\usepackage{comment}
\usepackage{wrapfig,lipsum,booktabs}
\usepackage{color}
\usepackage{url}
\usepackage{bm}
\usepackage{amsmath}
\usepackage{amsfonts}
\usepackage{mathrsfs}
\usepackage{wrapfig}
\usepackage{xcolor}
\usepackage{hyperref}
\usepackage[normalem]{ulem}
\newcommand{\beq}{\begin{equation}}
\newcommand{\eeq}{\end{equation}}
\newcommand{\beqa}{\begin{eqnarray}}
\newcommand{\eeqa}{\end{eqnarray}}

\newcommand{\df}{\text{distribution function}}
\usepackage{todonotes}
\DeclareMathOperator*{\argmin}{arg\,min}

\begin{document}

\input{Title/title}


\input{Abstract/abstract}

\input{Introduction/introduction}

\input{Numerical_method/numerical_method}

\input{Inference_to_solveVlasov/Inference}

\input{SolvePoisson_to_IPINN/solvePoisson}

\input{SolvePoisson_to_IPINN/IPINN}

\input{Conclusion/conclusion}

\section*{Acknowledgments}
This work has received financial support from the AIM4EP project
(ANR-21-CE30-0018), funded by the French National Research Agency (ANR) and was granted access to the HPC resources of IDRIS under the allocations 2021-AD011011719R1 and 2022-AD011011719R2 made by GENCI.
The authors are deeply thankful for the financial assistance provided by A*MIDEX for ISFIN-SRIM research mobility for PhD students. Their support enabled us to carry out the research work with our collaborators at Jülich Supercomputing Centre in Forschungszentrum Jülich. The authors gratefully acknowledge computational resources provided by the Gauss Centre for Supercomputing e.V. (www.gauss-centre.eu) on the GCS Supercomputer JUWELS \cite{JUWELS} at Jülich Supercomputing Centre (JSC).
\section*{References}
\bibliographystyle{unsrt}
\bibliography{reference}

\appendix
\input{Appendix/appendix}

\end{document}

%% file: Title/title.tex
\title[PINN applied to wave-particle resonance in fusion plasmas]{Physics informed Neural Networks applied to the description of wave-particle resonance in kinetic simulations of fusion plasmas}

\author{Jai Kumar$^{1,2}$, David Zarzoso$^2$, Virginie Grandgirard$^1$, Jan Ebert$^3$, Stefan Kesselheim$^3$}
\address{$^1$ CEA, IRFM, F-13108 Saint-Paul-lez-Durance, France.}
\address{$^2$ Aix-Marseille Université, CNRS, Centrale Marseille, M2P2 UMR7340, Marseille, France.}
\address{$^3$ Jülich Supercomputing Center, Forschungszentrum Jülich, Wilhelm-Johnen-Strasse, 52428 Jülich, Germany.}
\ead{jai.kumar@cea.fr}
\vspace{10pt}

%% file: Abstract/abstract.tex
\begin{abstract}
The Vlasov-Poisson system is employed in its reduced form version (1D1V) as a test bed for the applicability of Physics Informed Neural Network (PINN) to the wave-particle resonance. Two examples are explored: the Landau damping and the bump-on-tail instability. PINN is first tested as a compression method for the solution of the Vlasov-Poisson system and compared to the standard neural networks. Second, the application of PINN to solving the Vlasov-Poisson system is also presented with the special emphasis on the integral part, which motivates the implementation of a PINN variant, called Integrable PINN (I-PINN), based on the automatic-differentiation to solve the partial differential equation and on the automatic-integration to solve the integral equation.
\end{abstract}

%% file: Introduction/introduction.tex
\section{Introduction}
Nuclear fusion reveals as the most promising solution for the increasing energy demand. However, a fusion plasma is an extremely complex system, characterized by large gradients of density, pressure and current that lead to the triggering of instabilities. Among those instabilities one may find those due to temperature and density gradients resulting in turbulent transport \cite{guzdar1983ion, lee1987collisionless, guo1989ion, cowley1991considerations, dorland2000electron, jenko2000electron, dong2002electron}, those driven by the energetic particles (such as the alpha particles produced by the fusion reactions) \cite{chen2016physics, lauber2013super}, or those driven by the gradients of the magnetic equilibrium \cite{furth1963finite,waelbroeck2009theory}. In general, such instabilities are deleterious for the overall confinement of particles and energy. This is the reason why understanding, predicting and eventually controlling instabilities in burning nuclear fusion plasmas is of prime importance on the route towards the steady-state production of energy in future fusion devices such as ITER or DEMO.

The study of instabilities can be done by means of analytic theory, experimental measurements or numerical simulations. One of the most complete frameworks for numerical analyses relies upon the kinetic description of the plasma, which requires solving the Vlasov equation (or Boltzmann equation in the presence of collisions) coupled to the Maxwell equations (Poisson equation and Ampère's law). Such approach involves numerical treatment of equations in 6D phase-space, which is numerically expensive and sometimes unaffordable. An approximation consists in averaging out the fastest gyromotion (cyclotron motion), whose frequency is usually much larger than the characteristic frequencies of the instabilities in fusion plasmas. Such approach is the fundamental basis of the so-called gyrokinetic codes \cite{brizard2007foundations}, which has been widely used for turbulent transport studies \cite{garbet2010gyrokinetic}, energetic particle physics \cite{zarzoso2013impact} and macro-scale instabilities linked to the magnetic equilibrium \cite{hornsby2015linear}. The advantage of the gyrokinetic approach is that the dimension of the problem is reduced down to 5D. Nonetheless, the high-dimensionality of the data implies that the cost of gyrokinetic simulations is still quite high, in terms of both computational time required to perform the simulations and disk space required to store the data to be post-processed. For this reason, new techniques to either accelerate the gyrokinetic simulations or optimize the way data is stored are mandatory.

In this framework, the use of artificial intelligence (AI) techniques might provide solutions to deal with high-dimensionality data from gyrokinetic simulations. In this paper, we explore the possibility of using deep neural networks to reduce the percentage of data that is required to be stored. Such approach might also be used online as a real-time processing of data and, in the limit of no stored data, results in solving the integro-differential equations by means of only neural networks.

Deep Learning has evolved since the 1990s and transformed how research is done in bio-informatics \cite{ai_protein}, self-driving cars \cite{self_drive_ai}, natural language processing \cite{nlp_ai}, and further applications \cite{baldi2014searching,dieleman2015rotation}. The usual application of deep learning is through the data-driven paradigm. In other words, large amount of data whose underlying model is unknown can be used to derive a reduced model based on artificial neural networks organized in sequential layers. The use of sequential layers (hence the designation of \textit{deep learning}) helps capture non-linearities in the data and augments the capability of representation. Such reduced model is not analytical and can be only applied if one has the weights and biases of the neural network. A step forward can be taken when dealing with data from a physical system whose evolution is given by a set of equations. Coupling the data-driven paradigm with the information from the equations can increase the capability of representation with a higher degree of accuracy with respect to the case where no information from the equation is employed \cite{NN_for_ODE}. In that context, Physics informed Neural Networks (PINN) \cite{raissi2017physics} are one class of neural networks exploiting the capability of deep neural networks as universal approximators \cite{HORNIK1989359} and encoding the differential equations underlying the physical system with the help of automatic differentiation \cite{baydin2018automatic}.

The natural question that arises when using PINN, is "why do we need neural networks if we already have the equations that govern the evolution of the physical system?". Indeed, solving the differential equations by means of standard numerical methods such as finite differences or finite elements might be more appropriate. However, using these methods to get the solution requires storing the numerical solution on a large number of grid points with a certain temporal resolution. For realistic physical systems this approach implies storing large amounts of data. In that respect, storing a neural network that can be used to obtain the solution at any point might represent a solution to optimise the storage. Depending on the amount of data used to train the neural network, PINN can exhibit different applications. For instance, it can be applied to infer missing data from known (or stored) data. When reducing the amount of known data down to the points at initial and boundary conditions, PINN is nothing else but a new way of solving differential equations. In the context of solving differential equations, PINN exhibit the advantage of being mesh-independent. Whereas PINN has already been used to solve linear and non-linear Partial Differential Equations (PDEs) \cite{raissi2017physics,Raissi2017PhysicsID,raissi2019physics}, there are some variants of PINN used to solve integro-differential equations \cite{f-PINNS, A-PINN, lindell2021autoint, kumar2022integral}, upon which part of our work relies.

In this paper we apply PINN to a specific physical problem: the wave-particle resonance \cite{escande2018basic}. Such mechanism is at the origin of most of the instabilities in fusion plasmas. Indeed, for an instability to occur, energy must flow from the particles of the plasma to the electromagnetic field. This exchange of energy is only possible when the phase velocity of the waves is close to the velocity of particles, which is the wave-particle resonance. Particles resonating with the wave but slightly faster (resp. slower) than the wave will transfer (resp. receive) energy to (resp. from) the wave. If on average there are more particles providing energy to the wave than particles receiving energy from the wave, the amplitude of the wave increases and an instability is triggered. This is the fundamental mechanism of the bump-on-tail instability \cite{bot}. When the opposite happens, the wave is damped. This is the mechanism for the Landau damping \cite{landau}. In order to simulate the two mechanisms we need to solve a system of coupled integro-differential equations, known as Vlasov-Poisson system. The Vlasov equation is a PDE and the Poisson equation is an integro-differential equation (IDE). We will use the VOICE code \cite{BOURNE2023112229} to numerically solve this system of coupled equations. We use the numerical results as ground truth for training AI models and also to compare the predictions after training. Since both mechanisms are ubiquitous in burning fusion plasmas, we focus on the present paper on the application of AI to capture the underlying physics by means of PINN-based models. In particular, we will explore three applications of PINN to the wave-particle resonance. 

\begin{itemize}
    \item[1.] Compare the use of Deep Neural Network (DeepNN) and PINN for storing the simulation data by using small percentage of stored results for training and inferring the rest.
    \item[2.] Use PINN to solve the Vlasov equation (the PDE in our system) and use the stored result of the Poisson equation (the IDE in our system) to predict the solution of coupled equation.
    \item[3.] A variant of PIIN, called I-PINN (for Integrable-PINN) is implemented to solve the integro-differential equation by combining automatic differentiation and the fundamental theorem of Calculus, inspired by \cite{lindell2021autoint} and \cite{kumar2022integral}. We also compare the results provided by I-PINN with the already existing f-PINN method \cite{f-PINNS}.
\end{itemize}
The remainder of the paper is structured as follows. Section \ref{NM} is devoted to introducing numerical methods used in the VOICE code to solve the Vlasov-Poisson system. The PINN method together with the employed architecture are also briefly discussed.
Section \ref{SSSP} is devoted to hyper-parameter tuning of DeepNN and PINN in case of inferring the missing data using very small amount of stored results. Both PINN and DeepNN are compared for the inference of missing data using least amount of stored data. At the end of section \ref{SSSP}, PINN is used to solve the Vlasov equation. In section \ref{fpinn_ide} f-PINN is used for the first time to solve the Vlasov-Poisson system using only the boundary and initial conditions. Finally, the I-PINN method is introduced in section \ref{ipinn_ide} to solve integro-differential equations and applied to the Vlasov-Poisson system. Section \ref{conclusion_futurework} concludes the paper and summarizes our work and future plans.

%% file: Numerical_method/numerical_method.tex
\section{Numerical methods: VOICE and PINN}\label{NM}
In the context of gyro-kinetic simulations for fusion plasmas, GYSELA exhibits the advantage that it is global, full-f and flux-driven \cite{Grandgirard_CPC2016}. It simulates the electrostatic plasma turbulence and induced transport in the core of the tokamak, evolving the 5-dimensional (3D in real space coordinates, and 2D in velocity coordinates) guiding-center distribution function of ions and electrons. Owing to its complexity, a reduced version of GYSELA has been developed, called VOICE \cite{BOURNE2023112229}. It is a two-dimensional (1D in real space, and 1D in velocity space) kinetic code. In this paper, before applying PINN to GYSELA, we explore its applicability to the VOICE code, which we use to simulate the Landau damping and the bump-on tail instability. These two physical mechanisms are used as test-bed cases for exploring the use of PINN to capture the physics of the wave-particle resonance.
In the following subsections we discuss briefly the VOICE code, the results for the Landau damping and the bump-on-tail instabilities as well as the PINN method, with special emphasis on the loss function and the training details.
\input{Numerical_method/voice}
\input{Numerical_method/pinn}
\input{Numerical_method/pinn_applied_to_voice}

%% file: Numerical_method/voice.tex
\subsection{VOICE to the solve the Vlasov-Poisson system}\label{voice}
The Vlasov-Poisson system describes the collective behaviour of charged particles in plasmas. It combines the Vlasov equation, which describes the kinetics of particles, with the Poisson equation, which relates the electric potential to the charge distribution. The Vlasov equation governs the evolution of the distribution function $f(x, v, t)$, which represents the distribution of particles at a given position $x$ with velocity $v$ at a time $t$. In the following, we assume a plasma with electrons and ions. Electrons are much faster than ions, therefore we assume on the analyzed time scale that the ions are at rest, so the only distribution function to obtain is that of electrons. The normalised Vlasov equation is given by
\begin{equation}
\label{Vlasov_eq}
   \frac{\partial f(x,v,t)}{\partial t}+v\frac{\partial f(x,v,t)}{\partial x}+ E(x,t)\frac{\partial f(x,v,t)}{\partial v}=0,
\end{equation}
where, $E(x,t)$ is the electric field, derived from the electric potential $E\left(x,t\right)=-\partial_x\phi\left(x,t\right)$. The evolution of the electric potential is given by the normalised Poisson equation, which expresses the relation between the potential and the charge density in the plasma
\begin{equation}
\label{Poisson_eq}
    -\frac{\partial^2 \phi(x,t)}{\partial x^2}= \int f(x,v,t)dv - 1.\\
\end{equation}
The VOICE code computes the solutions for $x\in\left[x_i,x_f\right]$ and $v\in\left[v_i,v_f\right]$ and restricting $t$ to the interval $\left[0,t_f\right]$.
For a detailed explanation of the normalization of equations \ref{Vlasov_eq} and \ref{Poisson_eq} as well as the numerical method employed in VOICE, the reader is encouraged to refer to \cite{BOURNE2023112229}. We impose Neumann boundary conditions in velocity for the distribution function,
\beq
\left.\frac{\partial f}{\partial v}\right\vert_{v=v_i}=\left.\frac{\partial f}{\partial v}\right\vert_{v=v_f}=0,\qquad\forall\left(x,t\right)
\label{nbc}
\eeq
Furthermore, we impose periodic boundary conditions in $x$ for both the distribution function and electrostatic potential,
\beqa
f\left(x_i,v,t\right) &=& f\left(x_f,v,t\right),\qquad\forall\left(v,t\right)\\
\phi\left(x_i,t\right) &=& \phi\left(x_f,t\right),\qquad\forall t
\label{pbc}
\eeqa
The initial conditions are simply $f\left(x,v,t=0\right)=f_0\left(x,v\right)$ and $\phi\left(x,t=0\right)$ is obtained from the Poisson equation \ref{Poisson_eq} using $f_0$. The parameters used for each of the VOICE simulations reported in this paper are summarized in table \ref{sp}. The discretization in $\left(x,v\right)$ is given by $\left (N_x, N_v\right)$, the number of points used in the respective dimensions.
\begin{table}[h!]
    \centering
    \begin{tabular}[width=15cm]{clp{1cm}cccccp{1.5cm}}
    \toprule 
    
    {Case}& {Simulation} & {$\varepsilon,\,\alpha$} & {$\Delta t$} & {$(N_x,N_v)$} & {$\left[0,t_f\right]$} & {$\left[x_i,x_f\right]$} & {$\left[v_i,v_f\right]$} & {Storage (MB)} \\\midrule
         
    I & LLD & $\varepsilon$=0.01 & 0.0125 & $(128,512)$ & [0,45] & $[0, 4\pi]$ & $[-6,6]$ & 1804 \\\midrule
         
    II & NLLD & $\varepsilon$=0.1 & 0.0125 & $(128,512)$ & [0,45] & $[0, 4\pi]$ & $[-6,6]$ & 1804 \\\midrule 
         
    III & BOT & $\alpha$=0.1 & 0.05 & (256,1024) & $[0,100]$ & $[0, 50]$ & $[-8,8]$ & 4200 \\\bottomrule
    
    \end{tabular}
    \caption{Simulation parameters of the three cases considered in this paper. LLD stands for \textit{linear Landau damping}, NLLD stands for \textit{non-linear Landau damping} and BOT stands for \textit{bump-on-tail}. $\Delta t$ is the time step, $N_x$ and $N_v$ are the total number of equidistant points chosen to solve equations \ref{Vlasov_eq} and \ref{Poisson_eq} in $x$ and $v$, respectively. Storage is the disk-space used to store the results of each simulation, namely the distribution function values at each point of $(x,v,t)$ and the electric potential at each $(x,t)$.}
    \label{sp}
\end{table}
\par In this article, we explore two cases: Landau damping and the bump-on-tail instability, as mentioned earlier. Without the loss of generality, the initial distribution function $f_0$ is decomposed into equilibrium ($f_{eq}$) and perturbed ($\delta f$) distribution functions. In our case, the initial perturbation is simply proportional to $f_{eq}$, as $\delta f = f_{eq}\varepsilon\cos{kx}$. Depending on the choice of $f_{eq}$, we will analyse either the Landau damping or the bump-on-tail instability. The phenomenon of Landau damping will be explored with the following equilibrium distribution function,
\beqa
\label{eq_f0}
f_{eq} = \frac{1}{\sqrt{2\pi}}\exp\left(\frac{-v^2}{2}\right).
\eeqa
\par
\begin{figure}
    \centering
    \includegraphics[width=7.5cm]{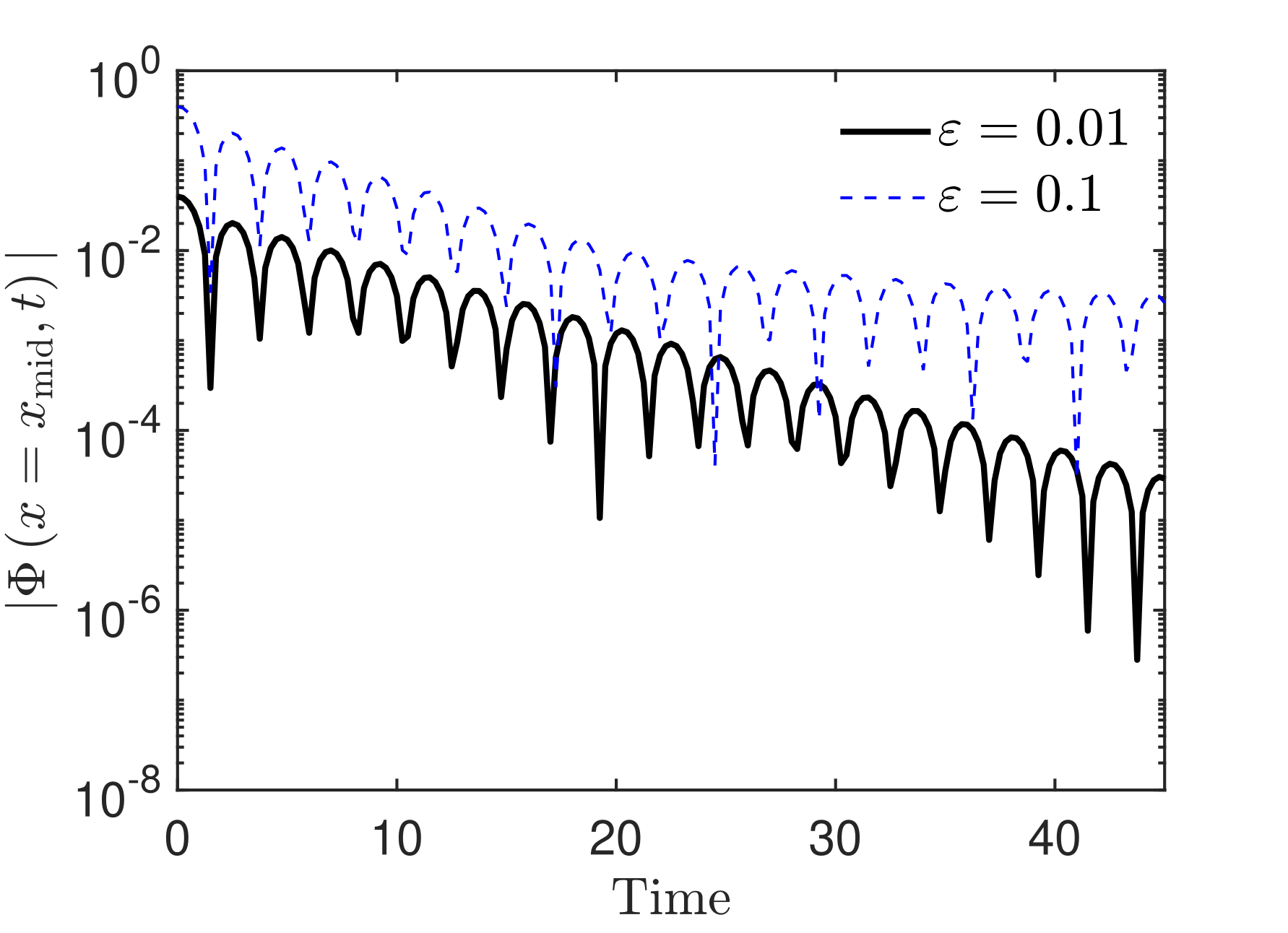}
    \includegraphics[width=7.5cm]{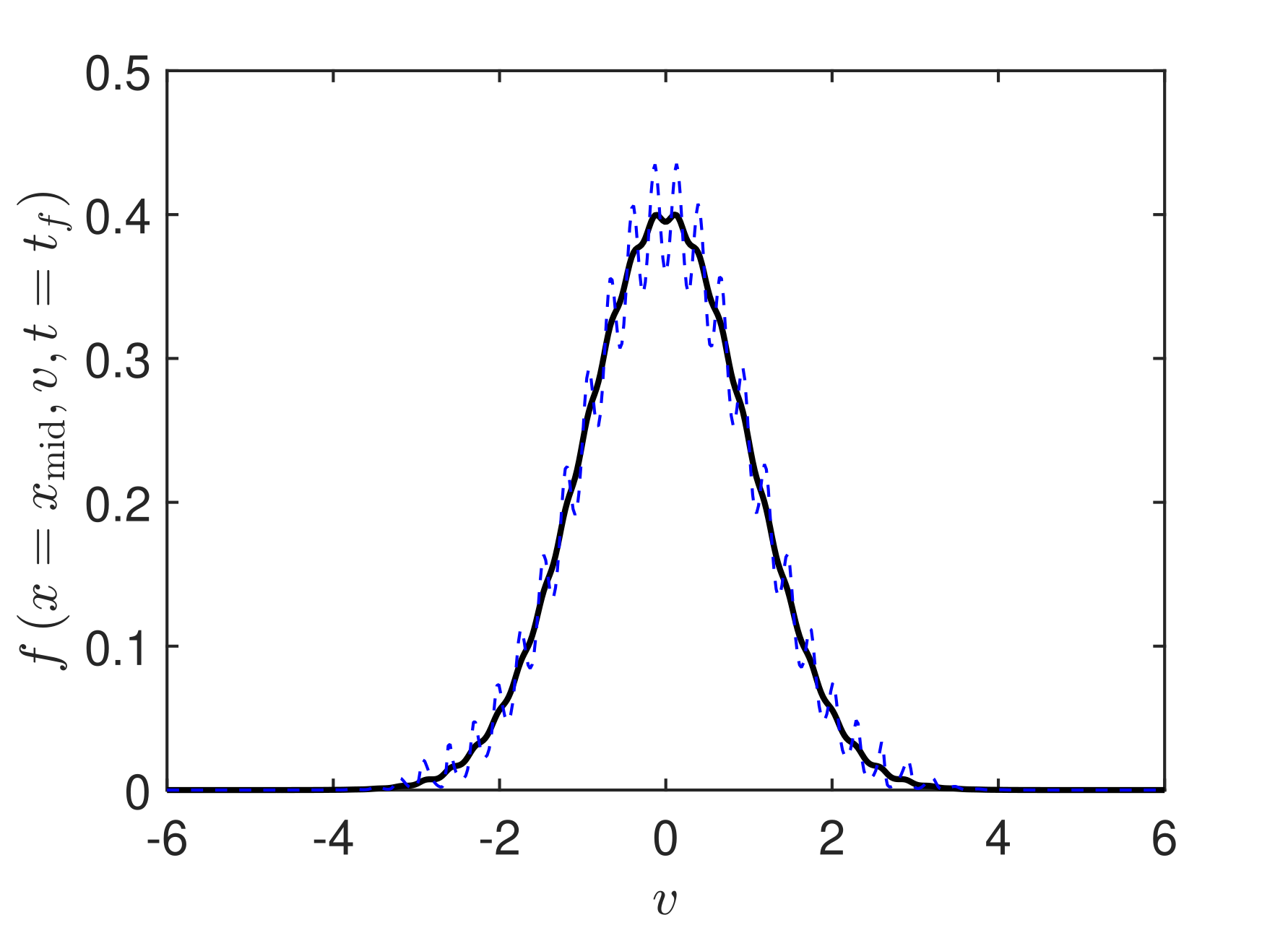}\\
    \includegraphics[width=7.5cm]{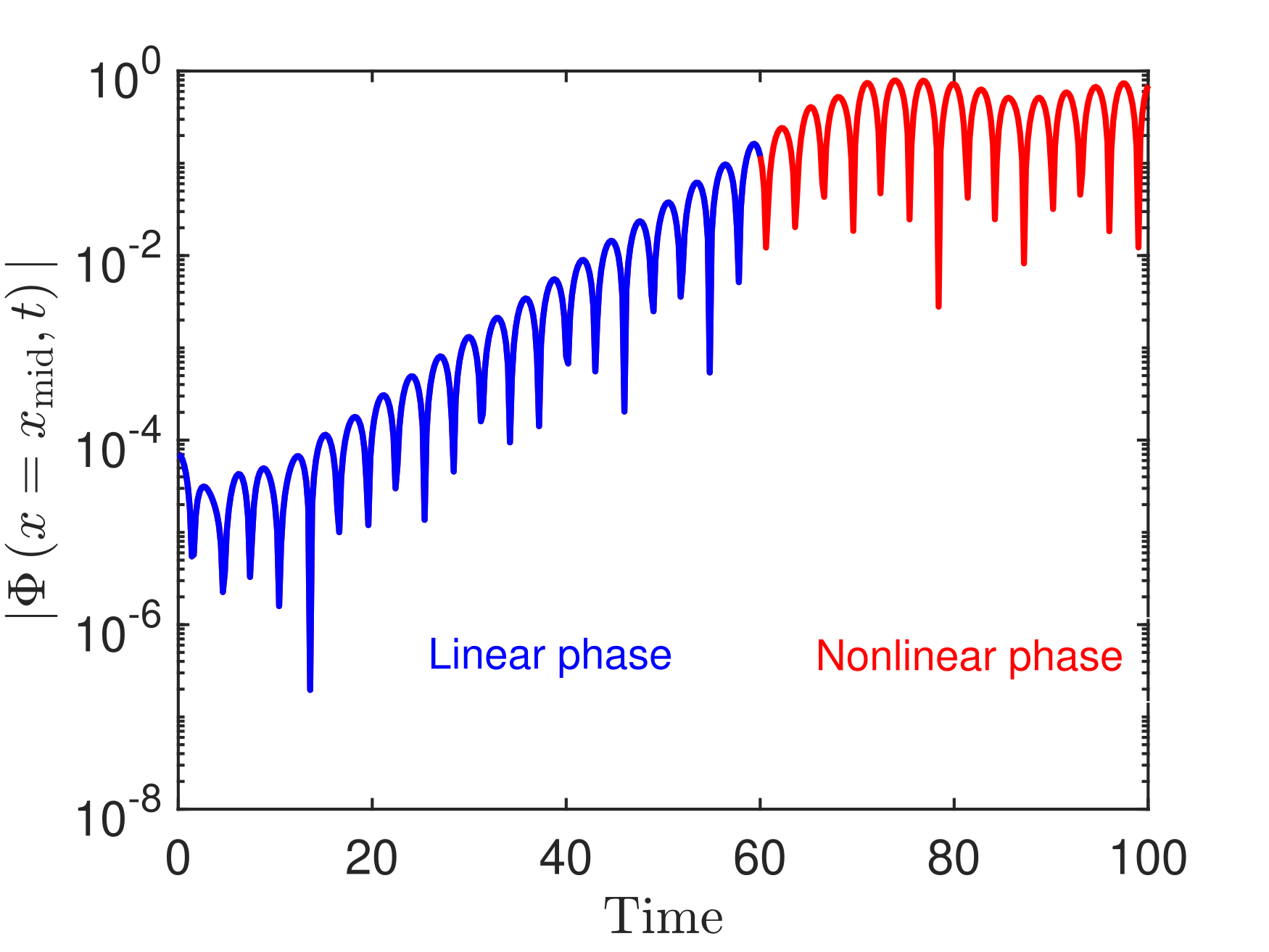}\label{Phi_vs_time_BOT_VOICEXX}
    \includegraphics[width=7.5cm]{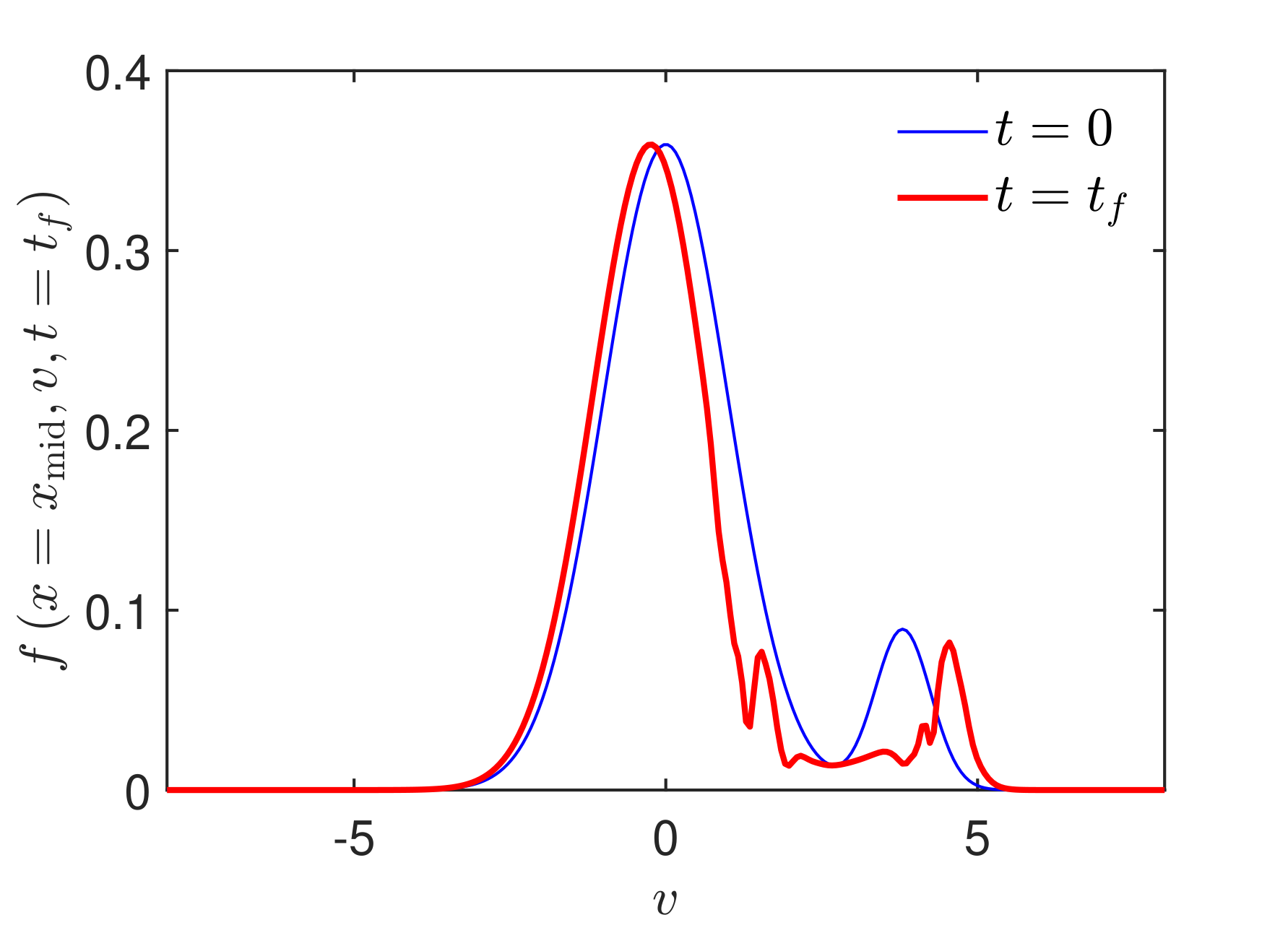}
    \caption{Summary of physical results from VOICE simulations. Top panels: time evolution of the electrostatic potential (left) evaluated at the mid $x$-position for the two considered Landau damping cases: $\varepsilon=0.01$ (thick solid black line) and $\varepsilon=0.1$ (thin dashed blue line). The final distribution function is evaluated at the mid $x$-position as a function of the velocity (right) for the two considered cases of Landau damping. Bottom panels: same quantities, but for the bump-on-tail instability. For the electrostatic potential, the linear and nonlinear phases are highlighted by thick blue and red lines, respectively. The final distribution function on the right figure is represented by a thick red line, together with the initial distribution represented by a thin blue line, where the bump-on-tail is clearly seen.}
    \label{Phi_fdist_Landau_bot}
\end{figure}

 VOICE has been run for this case up to $t_f$=45. Two values for the initial perturbation, $\varepsilon=0.01$ and $\varepsilon=0.1$, are employed for same $k=0.5$. Indeed, using the amplitude $\varepsilon=0.01$ leads to a linear damping during the whole simulation, whereas the amplitude $\varepsilon=0.1$ results in a linear phase followed by a nonlinear phase where the damping rate is modified. This nonlinearity leads to higher amplitude oscillations of the distribution function in phase space. The top panel of figure \ref{Phi_fdist_Landau_bot} summarizes the results obtained with VOICE for the Landau damping cases. The top right panel represents the distribution function for $\varepsilon=0.01$ (thick solid black line) and $\varepsilon=0.1$ (thin dashed blue line). The left panels represent the time evolution of the electrostatic potential evaluated at the mid $x$-position. The $y$-axis is given in logarithmic scale. It is observed that Landau damping is characterized by the formation of structures (or oscillations in velocity space). The amplitude of these oscillations increases with $\varepsilon$, but the location remains unchanged. The increasing amplitude is characteristic of the nonlinearities that play a major role in modifying the damping rate at $t\approx 25$.
\par Moving on to the bump-on-tail instability, analysis of the physics requires inverting the slope of the distribution function. Therefore, we consider an equilibrium distribution function $f_{eq}$ composed of a Maxwellian distribution $f_1$ with density $1-\alpha$ ($0\leq \alpha \leq1$), and a shifted Maxwellian distribution $f_2$ representing a population with a mean velocity $v_0$, density $\alpha$ and temperature $T_0$. The parameter $\alpha$ controls the fraction of super-thermal particles leading to an inversion of the slope of the total distribution function. The equilibrium distribution function therefore yields
\begin{align}
f_{eq}=&f_1+f_2,\,\rm{with}\\\nonumber
    &f_1 = \frac{1-\alpha}{\sqrt{2\pi}}\exp\left\{ \frac{-v^2}{2}\right\}\\ \nonumber
    &f_2 = \frac{\alpha}{\sqrt{2\pi T_0}}\exp\left\{ \frac{-(v-v_0)^2}{2T_0}\right\}
\end{align}
\label{eq_f_1_f2}
As for the Landau damping, the initial $\df$ is
\begin{equation}
    f_0=f_{eq}(1+\varepsilon\cos{kx})
\label{f_0_bot}
\end{equation}
This case has been simulated with VOICE till $t_f$=100. The parameters used are: $\varepsilon=10^{-5}$, $\alpha = 0.1$, $T_0 = 0.2$, $v_0 = 3.8$ and $k=0.5$. The results are summarised in the bottom panels of figure \ref{Phi_fdist_Landau_bot}. For the electrostatic potential, the two phases (linear and nonlinear) are conveniently highlighted by blue and red lines, respectively. The final distribution function is plotted in red in the bottom right panel, together with the initial distribution function plotted in blue. The initial distribution function exhibits a bump on the tail, responsible for the so-called bump-on-tail instability. Particles are redistributed in phase space during the nonlinear phase, as can be observed in the final distribution function. This redistribution leads to the saturation of the instability.

The computed solutions of the Vlasov-Poisson system for the two physical mechanisms (Landau damping and bump-on-tail) are stored. The distribution function, $f(x,v,t)$ is recorded for each spatial position $x$ and velocity $v$ at every time step during the numerical integration of the Vlasov equation. Additionally, the solution of the Poisson equation, representing the electrostatic potential $\phi(x,t)$, is stored for every spatial position and time. It should be noted that storing the results of a single VOICE simulation, corresponding to $\phi$ and $f$ requires approximately 1 to 4 GB of disk space, as indicated in the "Storage" column of table \ref{sp}. These recorded data serve as the ground truth for training deep learning models in subsequent analyses.
 

%% file: Numerical_method/pinn.tex
\subsection{Physics Informed Neural Networks}\label{PINN}
The property of PINN to simulate a mesh-free physical system without requiring much data has attracted the interest of a lot of researchers resulting in quite a few variants of PINN \cite{Kress2021AMP}. Among those, fractional-PINN (f-PINNs) \cite{f-PINNS} is used to solve integro-differential equations, variational-PINN (vPINN) \cite{Kharazmi2019VariationalPN} uses a test space for Legendre polynomials to reduce the training cost, hp-vPINN is an extended version of vPINN. Many more variants exist and the reader is encouraged to go through the referred bibliography \cite{Saha2020PhICNetPC,Zhang2020LearningIM,Zhong2022PIVAEPV,Sun2020NeuPDENN,Long2018PDENetLP}.
\par
\begin{figure}[h!]
    \centering
    \includegraphics[width=10.5cm]{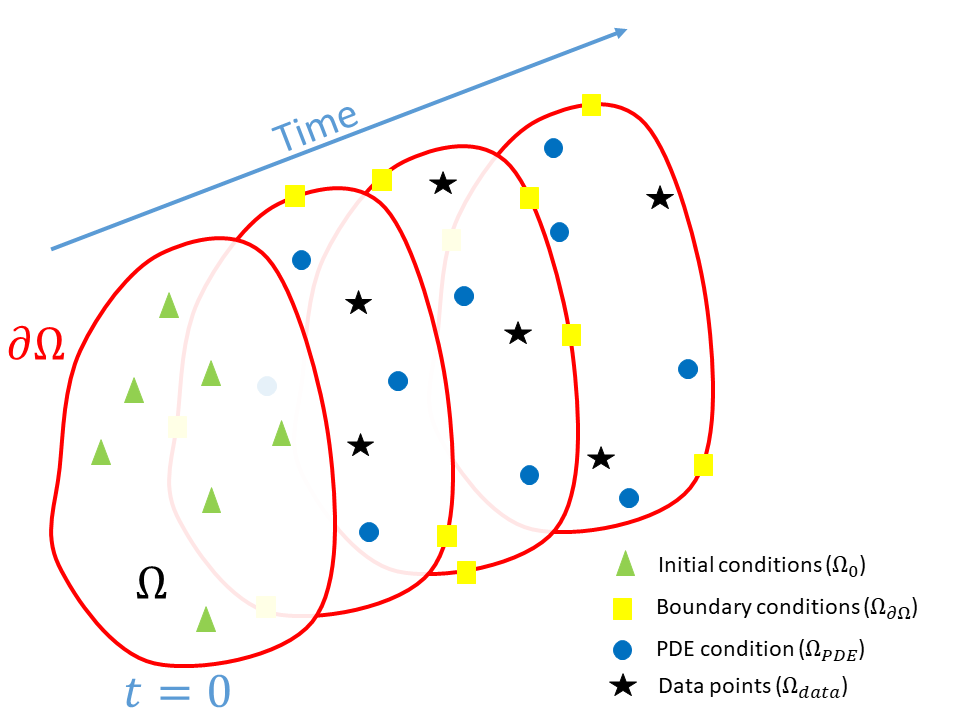}
    \caption{Schematic representation of the domain where the PDEs are defined as well as the subset of points where the neural networks are forced to minimize each term of the loss function. Here, we assume initial time, $t_{i} = 0$.}
    \label{fig-domain}
\end{figure}

 Mathematically speaking, PINN is a fully connected deep neural network that takes a vector $(\mathbf{y},t)$ in the defined phase-space of a PDE as input and returns as output an approximation to the solution of the PDE evaluated at that vector. The PDE is defined as a differential operator $\mathcal{L}$ applied to a function $f$ to give a function $g$
\begin{equation}
    \mathcal{L}\left(f\left(\mathbf{y},t\right)\right) = g(\mathbf{y},t), \qquad \left(\mathbf{y},t\right)\in\Omega\cup\left[\left.t_{i}, t_{f}\right]\right.,
\label{pde}
\end{equation}
where $\Omega$ is the sub-space on which the solution of the PDE is defined from initial time $t_{i}$ till final time $t_{f}$. The boundary of $\Omega$ is divided into $\partial\Omega=\partial\Omega_N\cup\partial\Omega_D\cup\partial\Omega_{P1}\cup\partial\Omega_{P2}$, where $\partial\Omega_N$ is the boundary where the solution satisfies Neumann conditions, $\partial\Omega_D$ is the boundary where Dirichlet conditions apply and $\partial\Omega_{P1}$ and $\partial\Omega_{P2}$ represent the boundaries where the solution satisfies periodic conditions. In addition, initial conditions must be satisfied. Therefore, the solution of the PDE is supposed to satisfy the following conditions
\beqa
f\left(\mathbf{y},t=t_{i}\right) &=& f_{0}\left(\mathbf{y}\right), \forall\mathbf{y}\in\Omega\\
\mathbf{n}\cdot\nabla f\left(\mathbf{y},t\right)&=& f_{\partial\Omega_N}\left(\mathbf{y},t\right), \forall\mathbf{y}\in\partial\Omega_N,\,\forall t \geq t_{i}\\
f\left(\mathbf{y},t\right)&=&f_{\partial\Omega_D}\left(\mathbf{y},t\right),\,\forall\mathbf{y}\in\partial\Omega_D,\,\forall t\geq t_{i}\\
f\left(\mathbf{y},t\right)&=&f\left(\mathbf{y}',t\right),\,\forall\left(\mathbf{y},\mathbf{y}'\right)\in\partial\Omega_{P1}\times\partial\Omega_{P2},\,\forall t \geq t_{i}
\eeqa
In the previous equation, $f_{0}$ is a function of $\mathbf{y}$ only, $\mathbf{n}$ is the normal vector to $\partial\Omega_N$ for the defined time range and $f_{\partial\Omega_N}$ and $f_{\partial\Omega_D}$ are functions of $\mathbf{y}$ and $t$.
The goal is to approximate the solution of the PDEs by neural networks, defined as $f_{NN}=f_{NN}\left(\mathbf{y},t;\boldsymbol{\theta}\right)$, where $\boldsymbol{\theta}$ represents the vector of weights and biases of the neural network and the subscript $NN$ stands for \textit{Neural Network}. To achieve this goal, an optimization problem is set up, aiming at minimizing a cost or a loss function $L$ made up of different terms, each term accounting for a given condition that the neural network must satisfy. These conditions are the following:
\begin{itemize}
    \item The network must satisfy the PDEs at any point.
    \item The network must satisfy the initial and boundary conditions.
    \item The network must be good approximator of the solution at the known or stored data points, referred to as \textit{data points} in the remainder of the paper.
\end{itemize}
In order to compute the derivatives of the neural network automatic differentiation is utilized, leveraging the chain rule to compute exact derivatives up to machine precision. Consequently, the overall expression of the loss function can be represented as the summation of the following terms:
\begin{equation}
    L = L_{\rm PDE} + L_{t=t_{i}} + L_{\partial\Omega_N} + L_{\partial\Omega_D} + L_{\partial\Omega_{P1}\times\partial\Omega_{P2}} + L_{\rm data}
    \label{loss}
\end{equation}
The losses are defined over all the points described before. In this paper, we use Mean Squared Error (\textit{MSE}) as loss function. In that case, each of these terms has the following expression
\begin{subequations}
\beqa
L_{\rm PDE}\left(\boldsymbol{\theta}\right)= \sum_{\mathbf{\hat{y}}\in\Omega_{{\rm PDE}}}\frac{\left\vert\mathcal{L}\left(f_{NN}\left(\mathbf{\hat{y}};\boldsymbol{\theta}\right)\right)-g\left(\mathbf{\hat{y}}\right)\right\vert^2}{N_{\rm PDE}} \\
L_{t=t_{i}}\left(\boldsymbol{\theta}\right)=\sum_{\mathbf{\hat{y}}\in\Omega_{0}}\frac{\left\vert f_{NN}\left(\mathbf{\hat{y}};\boldsymbol{\theta}\right)- f_{0}\left(\mathbf{\hat{y}}\right)\right\vert^2}{N_{0}} \\
L_{\partial\Omega_N}\left(\boldsymbol{\theta}\right)= \sum_{\mathbf{\hat{y}}\in\partial\Omega_{N}}\frac{\left\vert\mathbf{n}\cdot\nabla f_{NN}\left(\mathbf{\hat{y}};\boldsymbol{\theta}\right)-f_{\partial\Omega_N}\left(\mathbf{\hat{y}}\right)\right\vert^2}{N_{\rm \partial\Omega_{N}}}\label{LboundaryN} \\
L_{\partial\Omega_D}\left(\boldsymbol{\theta}\right)=\sum_{\mathbf{\hat{y}}\in\partial\Omega_{D}}\frac{\left\vert f_{NN}\left(\mathbf{\hat{y}};\boldsymbol{\theta}\right) - f_{\partial\Omega_D}\left(\mathbf{\hat{y}};\boldsymbol{\theta}\right)\right\vert^2}{N_{\rm \partial\Omega_{D}}}\label{LboundaryD} \\
L_{\partial\Omega_{P1}\times\partial\Omega_{P2}}\left(\boldsymbol{\theta}\right)=\sum_{\left(\mathbf{\hat{y}},\mathbf{\hat{y}}'\right)\in\partial\Omega_{P1}\times\partial\Omega_{P2}}\frac{\left\vert f_{NN}\left(\mathbf{\hat{y}};\boldsymbol{\theta}\right)-f_{NN}\left(\mathbf{\hat{y}}';\boldsymbol{\theta}\right)\right\vert^2}{N_{\rm \partial\Omega_{P}}}\label{LboundaryP} \\
L_{\rm data}\left(\boldsymbol{\theta}\right)=\sum_{\mathbf{\hat{y}}\in\Omega_{{\rm data}}}\frac{\left\vert f_{NN}\left(\mathbf{\hat{y}};\boldsymbol{\theta}\right)- f\left(\mathbf{\hat{y}}\right)\right\vert^2}{N_{{\rm data}}}
\eeqa
\end{subequations}
where we have written $\mathbf{\hat{y}}\equiv\left(\mathbf{y},t\right)$ for the sake of readability. $\Omega_{{\rm PDE}}$ represents the subset of points where the neural network is forced to satisfy the PDE. These points are usually called \textit{collocation points} in the literature. $\Omega_{0}$ is the subset of points where the neural network is forced to satisfy the initial conditions. $\partial\Omega_{N}$, $\partial\Omega_{D}$, $\partial\Omega_{P1}$ and $\partial\Omega_{P2}$ are the set of points where the boundary conditions must be satisfied. $\Omega_{{\rm data}}$ is the set of points where the neural network is forced to approximate the stored data. We have also defined $N_{\rm PDE}=\#\Omega_{{\rm PDE}}$, $N_{ 0}=\#\Omega_{0}$, $N_{\rm \partial\Omega_{N}}=\#\Omega_{\partial\Omega_{N}}$, $N_{\rm \partial\Omega_{D}}=\#\Omega_{\partial\Omega_{D}}$, $N_{\rm \partial\Omega_{P}}=\#\Omega_{\partial\Omega_{P1}} = \#\Omega_{\partial\Omega_{P2}}$ and $N_{{\rm data}}=\#\Omega_{{\rm data}}$. The different subsets of points are schematically represented in figure \ref{fig-domain}. It is to be noted that since the neural network $f_{NN}$ depends on $\mathbf{\hat{y}}$ and also on the vector of weights and biases $\boldsymbol{\theta}$, each term of the loss function depends on $\boldsymbol{\theta}$. Therefore, the general minimization problems reads
\beq
\boldsymbol{\theta}^{*} = \argmin_{\boldsymbol{\theta}}L\left(\boldsymbol{\theta}\right)
\eeq

This minimization problem can be solved using standard optimization techniques, such as gradient descent, stochastic gradient descent or extended versions with adaptive learning rates, such as Root Mean Squared Propagation (\textit{RMSP}) and adaptive moment estimation (\textit{Adam}), where the learning rate is adapted using averaged first (for RMSP) or second (for Adam) moments of the gradients.
 

%% file: Numerical_method/pinn_applied_to_voice.tex
\par
\begin{figure}[h]
    \centering
    \includegraphics[width=16cm]{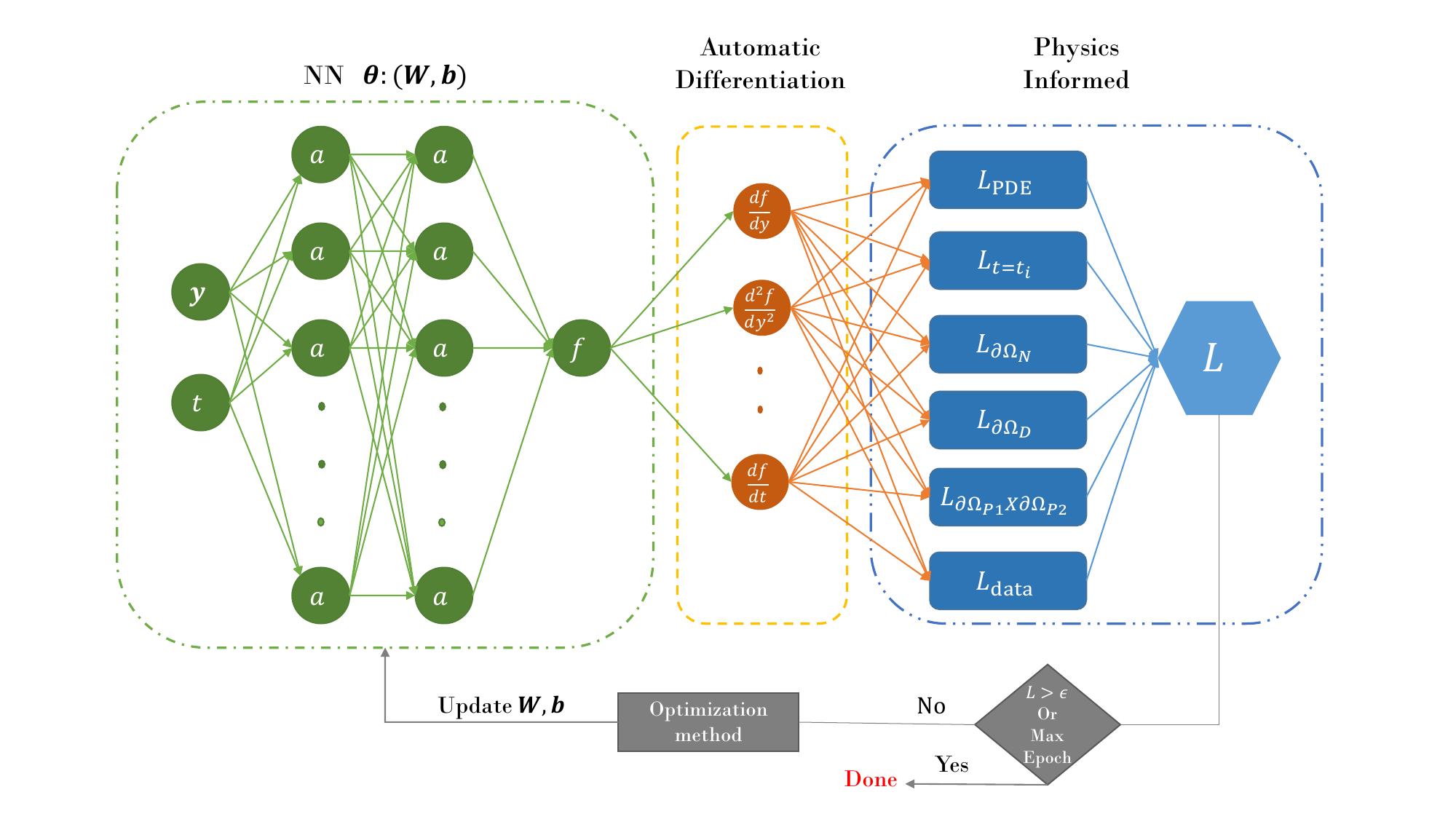}
    \caption{Schematic representation of the PINN workflow. A deep neural network is built with the activation function $a$. It takes $(\mathbf{y},t)$ as input and returns as output a value which is supposed to approximate $f$. The derivatives of the output $f$ are computed and evaluated numerically with respect to the inputs to find the loss for the desired PDE that we want to solve along with boundary conditions and initial conditions. The total loss is then computed by adding all the losses together. The training of the network is done using the total loss function until the desired accuracy or epochs are achieved.}
    \label{PINNworkflow}
\end{figure}

 The workflow to train PINN is represented schematically in figure \ref{PINNworkflow}. The neural network is composed of $N_{l}$ layers and each layer is composed of $N_{\rm n}$ neurons or nodes. Each node is characterized by an activation function $a$. The network takes as input vector a $\mathbf{y}$ and a time $t$. The network output is supposed to approximate the solution of the PDE, $f\left(\mathbf{y},t\right)$. The automatic-differentiation package computes the partial derivatives needed to calculate $L_{\rm PDE}$ and $L_{\partial\Omega_{N}}$ with respect to $\left(\mathbf{y},t\right)$. Similarly, other terms of the loss function ($L_{t=t_{i}}, L_{\partial\Omega_D}, L_{\partial\Omega_{P1}\times\partial\Omega_{P2}},  L_{\rm data}$) can be evaluated for the respective input values and required derivatives. All the terms are then summed up to evaluate the total loss.If the value of the loss function is smaller than the desire value or maximum iterations are reached, we consider the training to be finished. Otherwise, the weights and biases are re-adapted following the optimizer. The dataset is split into training and test datasets. Usually in this article, the proportion used is 90-10 unless specified otherwise. The training dataset is distributed into $N_{\rm b}$ batches. All the trainings in this paper use $N_{\rm b}$=10, unless specified otherwise. The weights and biases of the neural network are updated after each batch and after each epoch. Once the network is trained, it is used to predict the values for unseen $\left(\mathbf{y},t\right)$ inputs and compared with the test data set to produce the test loss. It is obvious that one needs to store the weights and biases of the trained Neural Network instead of storing the entire distribution function at each  $\left(x,v,t\right)$. The disk space required to store the trained Neural Network represents orders of magnitude smaller than the storage given in the table \ref{sp}. This is our motivation behind using PINN as a storage solution for exascale simulations.

There exists already many packages mainly written in Python using \textit{TensorFlow} or \textit{Pytorch} backend, like SciANN \cite{SciANN} and DeepXDE \cite{deepxde}. In this paper, we made our own program to suit our needs as these packages are currently under development. In particular, we use \textit{TensorFlow 2.8} \cite{tensorflow2015-whitepaper} and \textit{TensorFlow probability 0.15} \cite{dillon2017tensorflow} in Python along with \textit{Adam} optimizer in \textit{Keras} for training. \textit{TensorFlow} is GPU compatible for fast training of AI models. We implement data parallelization with the help of \textit{horovod} \cite{horovod-tensorflow} in tensorflow.
\par Defining the loss function for the coupled Vlasov-Poisson system is not straight forward. The Poisson equation is an integro-differential equation for which auto-differentiation techniques do not apply a priory. For this reason, the remainder of the paper is split into two applications. First, we will focus on solving PDE and apply PINN to the Vlasov equation. We will use the values of the distribution function and electric potential for this application. It can be seen as an inference problem where some values of the distribution function are known and the other values are missing. This has a clear interest towards optimising the storage of values in high-dimensional plasma simulations. It is to be noted that when the percentage of known values of the distribution function tends to zero (only remaining known data of the initial and boundary conditions), the inference simply results in solving the Vlasov equation. The second application will focus on solving the coupled Vlasov-Poisson system for which other techniques different from the auto-differentiation are required to integrate the neural network that approximates the distribution function.

%% file: Inference_to_solveVlasov/Inference.tex
\section{From the inference of the distribution function to the full integration of the Vlasov equation: a first step towards a solution to the storage problem of exa-scale simulations}\label{SSSP}
Performing extensive kinetic or gyrokinetic simulations of fusion plasmas is a challenging task in terms of the complexity of the numerical methods and the storage of the solution, which is intrinsically high-dimensional and therefore requires large disk space. It is the case of the code GYSELA-X, used in simulations of electrostatic plasma turbulence and transport in the core of tokamaks. Codes like GYSELA-X output heavy files (up to 2 TB) at every time step, which makes it impossible to store all the results. Therefore, in general, a compromise must be reached and one is constrained to save only some reduced part of the output every few time steps. This solution can, of course, jeopardize the understanding of the physics underneath and is prone to missing some crucial physical behaviour. It is the case in particular for the physical applications of the Landau damping and the bump-on-tail instability that we deal with in the present work. Both of these applications are based on the wave-particle interaction, which is the primary mechanism at the origin of kinetic instabilities in plasmas. The wave-particle interaction allows energy and momentum exchange between particles and waves, where nonlinear effects lead to the formation of structures that are localized in phase space. Access to these structures is important for understanding the interaction between particles and waves.
\par However, storing only a minor percentage of data might result in misleading interpretations of the data. For this reason, in this section, the application of neural networks to the wave-particle interaction is presented. Section \ref{deppNN} and \ref{PINN_infer} is devoted to fixing hyper-parameters of neural network for training DeepNN and PINN in case of inferring the missing data using very small amount of stored results. In Section \ref{compare_deepNN_PINN} the results obtained for inference of stored results using least amount of data for DeepNN and PINN are compared. In section \ref{PINN_solve_Vlasov}, PINN is used to solve the Vlasov equation utilizing only the stored results of Poisson equation, without using any data of distribution function. Different methods to choose training points (specifically collocation points) for training PINNs are also defined and compared. All the training in this section is done for different subsets of datasets given in table \ref{sp}. The subsets are in terms of the initial and final time of the chosen dataset. The reason is to capture the physical meaning and trend of the data with the limited resources of training AI models.
\input{Inference_to_solveVlasov/NNandPINN_for_inference}

\input{Inference_to_solveVlasov/solveVlasov}

%% file: Inference_to_solveVlasov/NNandPINN_for_inference.tex
\subsection{DeepNN applied to infer the stored simulation data}
\label{deppNN}
DeepNN has the same architecture as given in figure \ref{PINNworkflow}, the only difference comes from the loss function. The loss function $L$ used to train DeepNN is composed of just two terms, $L_{t=t_i}$ and $L_{data}$, as defined in equation \ref{loss}. In this case, 
\begin{align}
    L = & L_{t=t_{i}}\left(\boldsymbol{\theta}\right) + L_{\rm data} \left(\boldsymbol{\theta}\right)\\\nonumber
\text{where,}\\\nonumber
    & L_{t=t_{i}}\left(\boldsymbol{\theta}\right)=\sum_{\left(x,v,t_i\right)\in\Omega_{0}}\frac{\big | f_{DeepNN}\left(x,v,t_i;\mathbf{\boldsymbol{\theta}}\right)- f\left(x,v,t_i\right) \big|^2}{N_{0}} \\ \nonumber
    & L_{\rm data}\left(\boldsymbol{\theta}\right)=\sum_{\left(x,v,t\right)\in\Omega_{\rm data}}\frac{\left\vert f_{DeepNN}\left(x,v,t;\mathbf{\boldsymbol{\theta}}\right)- f\left(x,v,t\right)\right\vert^2}{N_{\rm data}} \nonumber
    \label{loss_deepNN}
\end{align}
where the output from DeepNN is denoted as $f_{DeepNN}$. For the training process, a small portion of the stored distribution function results are used. Specifically, we select a subset of Case-II data spanning from $t_i=0$ to $t_f=20$ to evaluate the performance of the DeepNN for inference. This subset is chosen due to the time required for the DeepNN to learn non-linear patterns within the data. By selecting a smaller dataset, we mitigate the training time and GPU resource requirements. However, despite not using the entire dataset, we can still capture the underlying physics of Landau damping effectively.
\par
\begin{figure}[h]
    \centering
    \subfloat[]{\includegraphics[width=7cm]{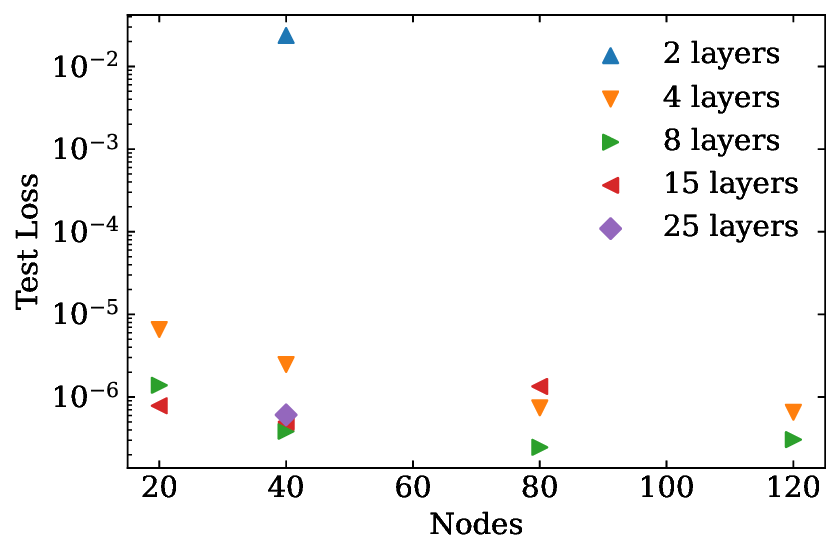}
    \label{NN_model_comparison_till_t20}}
  \subfloat[]{\includegraphics[width=7cm]{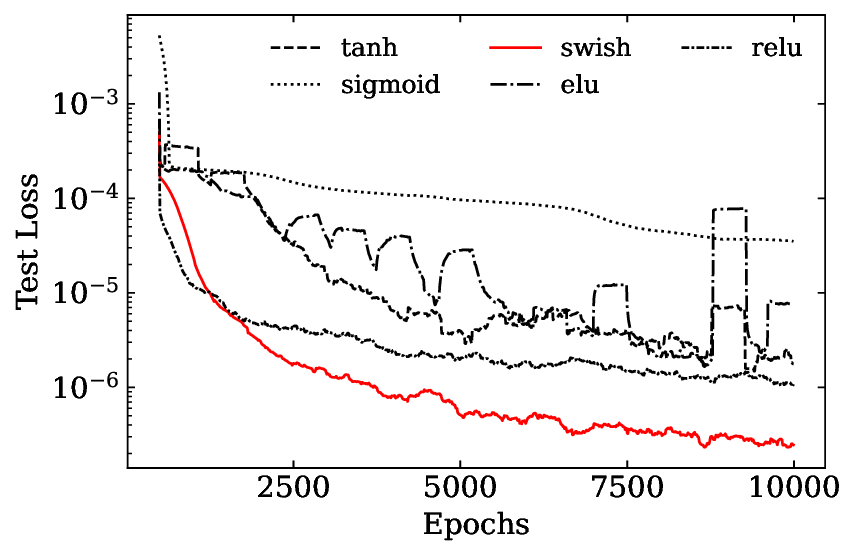}
    \label{PINN_model_comparison_till_t20}}
    \caption{(a) The final value of rolling average over 500 epochs of the test loss for 10000 epochs are plotted for different DeepNN architectures. The same colour and symbol data points represent the same number of layers with different number of nodes in each layer. The NN with 80 nodes in each of the 8 layers is trained for different activation functions, the test loss during training is plotted on the right (b). It can be observed that the training with \textit{swish} (shown in red solid line) converges faster than the other activation functions for the same number of epochs. }
    \label{model_comparision_nn}
\end{figure}

 We randomly choose 1\% ($\approx$9.2x$10^5$ points) of the stored distribution function data to estimate $L_{\rm data}$ and 80\% ($\approx$6.4x$10^4$ points) of the initial distribution function data to estimate $L_{t=t_i}$ in the training. The optimizer ($Adam$) and learning rate ($0.001$) are fixed for all the trainings. To determine the ideal configuration of the neural network, first we explore a range of options for the number of hidden layers and the number of nodes within each layer, keeping the activation function of each layer fixed to $swish$. Specifically, we vary the number of hidden layers from 2 to 25 and the number of nodes per layer from 20 to 120. By systematically testing different combinations within these ranges, we aim to identify the optimal configuration that yields the best performance. The final test losses after 10000 epochs of training with different DeepNN architectures are given in figure \ref{NN_model_comparison_till_t20}. The values for the test losses are actually computed by performing  rolling average over 500 epochs. This is done in order to smooth the trends of the loss as a function of the number of epochs. The lowest final test loss is observed in DeepNN with 80 nodes in each of the 8 layers. Neural networks deeper than 8 layers also do very well giving the final test loss to the single precision, of the order of $10^{-7}$. As expected, the training time increases exponentially as the NN gets deeper.
\par To determine the best activation function for the DeepNN with 80 nodes in each of the 8 layers, we try several activation functions. Some of the most common ones are the Rectified Linear Unit (\textit{ReLU}), the \textit{tanh}, the \textit{sigmoid}, the Scaled Exponential Linear Unit (\textit{SELU}), the Exponential Linear Unit (\textit{ELU}) and the \textit{swish}. The training curves of DeepNN using different activation functions for the best NN are given in figure \ref{PINN_model_comparison_till_t20}. It can be observed that the training with \textit{swish} (given in red solid line) converges faster than the other activation functions. According to the results, \textit{swish} activation function works best for the non-linear Landau damping data. Therefore, this is the one that will be used in all the trainings in the remainder of the paper.
\subsection{PINN applied to infer the stored simulation data}
\label{PINN_infer}
PINN is used here to infer the full distribution function using a small percentage of randomly chosen stored data. The architecture of the PINN follows the structure outlined in figure \ref{PINNworkflow}. In VOICE simulations, the boundary conditions are defined by equations \ref{nbc} and \ref{pbc}, which do not involve Dirichlet boundary conditions. Therefore, the total loss is modified by removing the $L_{\partial\Omega_D}$ term in expression \ref{loss},
\begin{equation}
    L = L_{\rm PDE} + L_{t=t_{i}} + L_{\partial\Omega_{N}} +  L_{\partial\Omega_{P1}\times\partial\Omega_{P2}} + L_{\rm data} 
    \label{loss_infer}
\end{equation}  
where,
\begin{align}
    L_{\rm PDE}\left(\boldsymbol{\theta}\right)=&\frac{1}{N_{\rm PDE}} \sum_{(x,v,t)\in\Omega_{{\rm PDE}}} \bigg | \frac{\partial f_{\rm PINN}\left(x,v,t;\boldsymbol{\theta}\right)}{\partial t}+ v \frac{\partial f_{\rm PINN}\left(x,v,t;\boldsymbol{\theta}\right)}{\partial x}+ \\\nonumber
    &E(x,t)\frac{\partial f_{\rm PINN}\left(x,v,t;\boldsymbol{\theta}\right)}{\partial v} \bigg |^2 \\\nonumber
    L_{t=t_{i}}\left(\boldsymbol{\theta}\right)=&\frac{1}{N_{0}}\sum_{(x,v,t)\in\Omega_{0}}\big | f_{\rm PINN}\left(x,v,t;\boldsymbol{\theta}\right)-f_{0}\left(x,v,t\right)\big |^2 \\\nonumber
    L_{\partial\Omega_D}\left(\boldsymbol{\theta}\right)= &\frac{1}{N_{\rm \partial\Omega_{D}}} \sum_{(x,v,t)\in\partial\Omega_{D}} \big | f_{\rm PINN}\left(x,v,t;\boldsymbol{\theta}\right) - f_{\partial\Omega_D}\left(x,v,t;\boldsymbol{\theta}\right)\big |^2\\\nonumber
    L_{\partial\Omega_N}\left(\boldsymbol{\theta}\right)= &\frac{1}{N_{\rm \partial\Omega_{N1}}} \sum_{(x,v_i,t)\in\partial\Omega_{N1}}\left\vert \frac{\partial f_{\rm PINN}\left(x,v,t;\boldsymbol{\theta}\right)}{\partial v}\right\vert^2 + \\\nonumber
    &\frac{1}{N_{\rm \partial\Omega_{N2}}} \sum_{(x,v_f,t)\in\partial\Omega_{N2}}\left\vert \frac{\partial f_{\rm PINN}\left(x,v,t;\boldsymbol{\theta}\right)}{\partial v}\right\vert^2 \\\nonumber
    L_{\partial\Omega_{P1}\times\partial\Omega_{P2}}\left(\boldsymbol{\theta}\right)=&\frac{1}{N_{\rm \partial\Omega_{P}}}\\\nonumber
    &\sum_{\left((x_i,v,t),(x_f,v,t)\right)\in\partial\Omega_{P1}\times\partial\Omega_{P2}}\big | f_{\rm PINN}\left(x_i,v,t;\boldsymbol{\theta}\right)-f_{\rm PINN}\left(x_f,v,t;\boldsymbol{\theta}\right)\big |^2 \\\nonumber
    L_{\rm data}\left(\boldsymbol{\theta}\right)=&\frac{1}{N_{\rm data}}\sum_{(x,v,t)\in\Omega_{\rm data}}\big | f_{\rm PINN}\left(x,v,t;\boldsymbol{\theta}\right)- f\left(x,v,t\right)\big |^2
\end{align}
The output from PINN is denoted as $f_{\rm PINN}$. The loss for Neumann boundary condition over $v$ is represented by $L_{\partial\Omega_N}$  on boundaries $\Omega_{N1}$ and $\Omega_{N2}$. The loss for periodic boundary conditions over $x$ is represented by $L_{\partial\Omega_{P1}\times\partial\Omega_{P2}}$ on boundaries $\Omega_{P1}$ and $\Omega_{P2}$. As explained earlier, vanilla PINN \cite{raissi2017physics} can only solve PDEs, so the Vlasov equation is used in the loss for PDE. This means that the stored result of Poisson equation, i.e. the electric potential, is required to evaluate $L_{PDE}$ for the collocation points. The stored electric potential data from the simulation is 2D, which is $N_v$ times smaller than the 3D distribution function data. So, even though we are using the full solution of Poisson equation, it represents a very small amount in terms of data storage as compared to the full distribution function.
\par Similar to the hyper-parameter tuning performed for the DeepNN, we also conduct hyper-parameter tuning for PINN. As already seen for DeepNN, training time increases exponentially as the network gets deeper. Using PINN implies that more gradients have to be evaluated which will take even more time and resources. Therefore, a subset of Case-II data from $t_i=15$ to $t_f=20$ is chosen to test the PINN for inference. We choose this subset out of the full dataset because of non-linearities at this time in simulation. Indeed, if PINN can correctly infer data for this time-interval, it can easily infer the data for all the previous times.
As before, 1\% of distribution function data is chosen to estimate $L_{data}$. Here, the points on different boundaries have to be chosen to evaluate the loss terms in equation \ref{loss_infer}. 80\% of stored values of input are randomly chosen for initial and boundary conditions, i.e., $\Omega_{0}$ ($\approx$5.2x$10^4$), $\partial\Omega_{N_1}$ ($\approx$4.1x$10^4$), $\partial\Omega_{N_2}$ ($\approx$4.1x$10^4$), $\partial\Omega_{P_1}$ ($\approx$1.6x$10^5$) and $\partial\Omega_{P_2}$ ($\approx$1.6x$10^5$). 5\% of collocation points are randomly chosen out of the mesh-grid of the stored input, i.e., $(x,v,t)$ ($\approx$1.1x$10^6$ points) over $\Omega_{\rm PDE}$ for $L_{\rm PDE}$. Also, the electric field (or electric potential) data-points used for evaluating the $L_{PDE}$ are $\approx$5.1x$10^4$.

\par
\begin{figure}[ht]
    \centering
    \subfloat[]{\includegraphics[width=7cm]{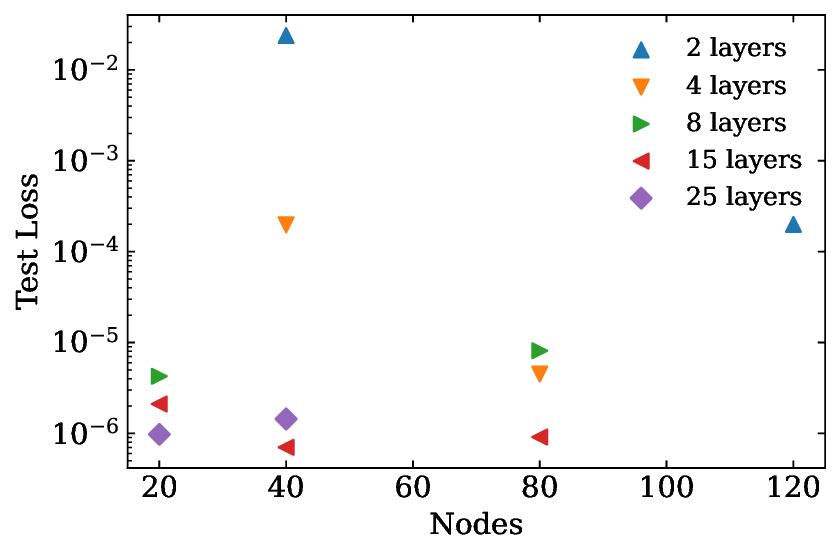}
    \label{PINN_model_comparison_from_t15_till_t20}}
  \subfloat[]{\includegraphics[width=7cm]{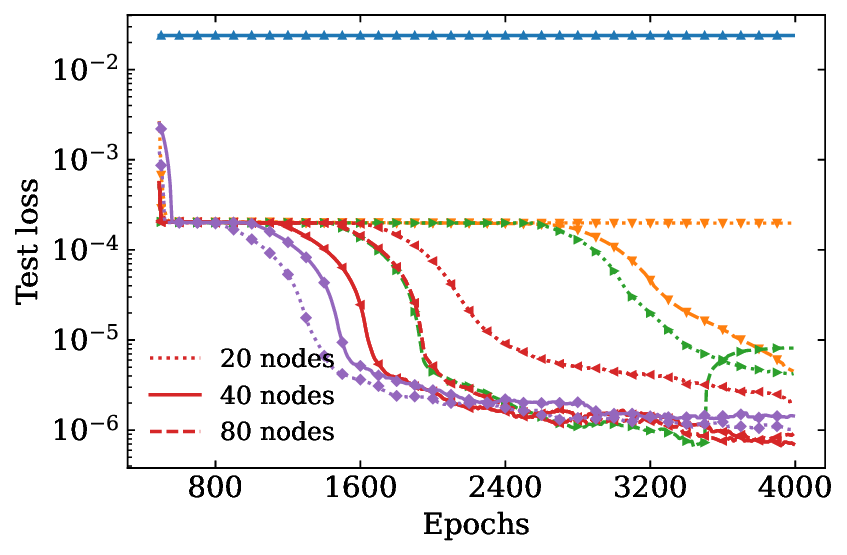}
    \label{PINN_model_comparison_losses_from_t15_till_t20}}
    \caption{(a) The final value of rolling average over 500 epochs of the test loss after 4000 epochs are plotted for different model architectures of PINN. The same colour data points represent the same number of layers with different numbers of nodes in each layer. The best model has 40 nodes in each of the 15 layers. (b) The test loss curves corresponding to the different model architectures are plotted with same color and symbol in the right with different line style for nodes. The same color and symbol represents the same number of layers, and the same line style represents the same number of nodes.}
    \label{model_comparision_pinn}
\end{figure}

 When training NN, evaluating the gradients is the most time consuming task. To train PINN, the loss $L_{PDE}$ is estimated by calculating three partial derivatives, which is not the case for DeepNN. This implies that the time required for PINN training is 3 to 4 times higher than that for DeepNN. Therefore, PINN is trained here up to 4000 epochs keeping all the training parameters the same as in section \ref{deppNN}. It can be seen from figure \ref{model_comparision_pinn} that the final test loss is of the order of $10^{-7}$ after 4000 epochs, which is small enough for an accurate prediction of the results.
Comparison of final test losses of different PINN architectures is given in figure \ref{PINN_model_comparison_from_t15_till_t20}. It can be seen from the figure that PINN needs to be deeper than 8 layers to give good prediction. The corresponding training curves are given in figure \ref{PINN_model_comparison_losses_from_t15_till_t20} showing that deeper PINN converges faster to lower test loss. A compromise has to be made between deeper PINN and training time. The best PINN architecture turns out to be 40 nodes in each of the 15 layers. It takes 5 GPU hours for the training to get a final test loss of the order of $10^{-7}$. One could argue that 8 layers with 80 neurons is the best model instead of 15 layers with 40 neurons. It is true that the former model converges faster for lower loss than the later but the training is unstable as the loss increases abruptly again. This abrupt change is mainly due to shallowness of the network. From now on, we only use NN with  40 neurons in each of 15 layers for our training with the \textit{swish} activation function. 

\begin{figure}[ht]
    \centering
    \includegraphics[clip, trim=4cm 4cm 10cm 2cm,width=16cm]{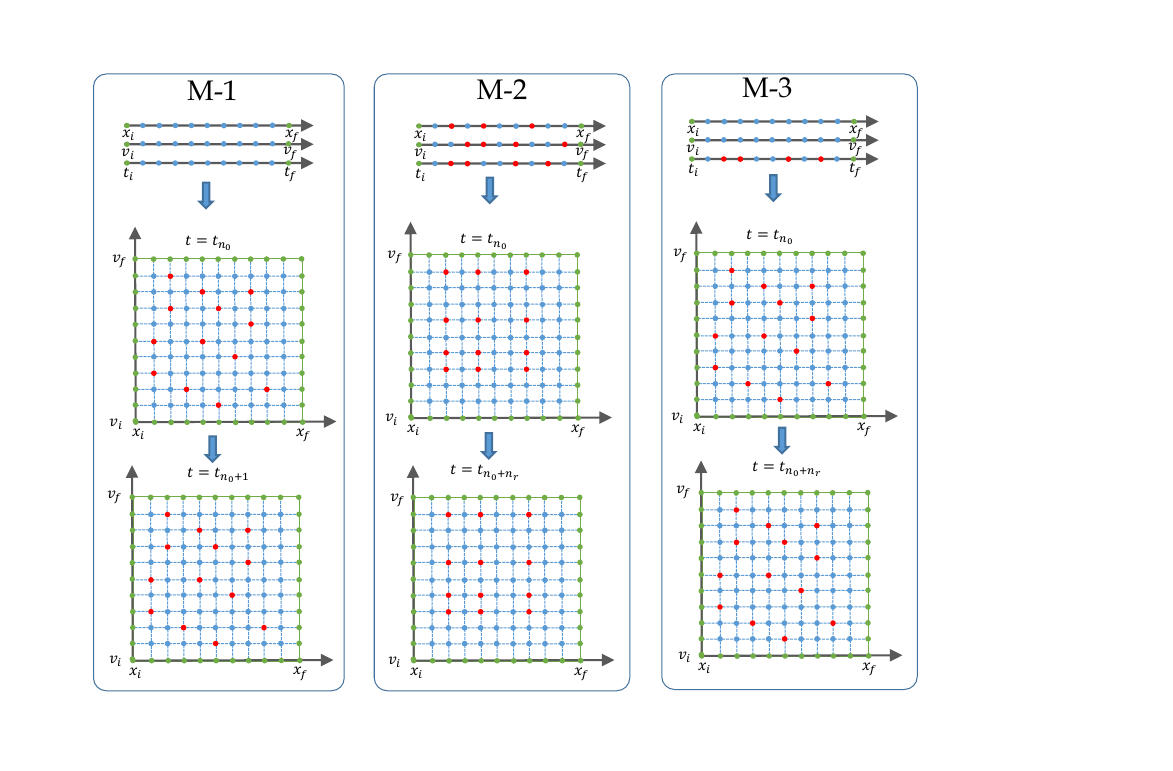}
    \caption{Schematic representation of different methods used to select training points corresponding to table \ref{table_methods}. Grid-points are represented in blue for $x$, $v$ and $t$ and the selected training points are represented in red. The selected points are shown for a given time represented by $t_{n_0}$. $t_{n_0+1}$ represents immediate time step after $t_{n_0}$ and $t_{n_0+n_r}$ represent next random time step after $t_{n_0}$ for which the training points are selected. M-1 selects points randomly from the mesh grid of $(x,v,t)$. Therefore, each time step will have some chosen training points. M-2 select random points directly from $x$-grid, $v$-grid and $t$-grid and make mesh grid of chosen points for training. M-3 selects random $t$-grid points and randomly chooses $(x,v)$ grid points for each chosen time steps.}
    \label{methodworkflow}
\end{figure}

Collocation points and distribution function data points were chosen randomly to evaluate $L_{PDE}$ and $L_{data}$ respectively, in the above trainings. Different ways can be employed for choosing these points to facilitate the training. Three methods to choose the training points are devised for our aim to store as less data from simulation as possible. These methods are summarised in table \ref{table_methods} and schematically represented in figure \ref{methodworkflow}.

\begin{table}[ht]
    \centering
    \begin{tabular}[width=15cm]{lp{10cm}}
    \toprule
    {Method}&{Description}\\
    \midrule
    M-1 & Randomly choose given percentage of points from the full phase space of $(x,v,t)$.\\
    M-2 & Randomly choose given percentage of points from $x$-grid, $v$-grid and $t$-grid separately and make a 3D mesh-grid using the chosen $(x,v,t)$ points for training.\\
    M-3 & Randomly choose given percentage of points from $t$-grid and for each of these timesteps randomly choose given percentage of points from 2D $(x,v)$ meshgrid.\\
    \bottomrule
    \end{tabular}
    \caption{Different methods to choose collocation points for $L_{\rm PDE}$ and distribution function points for $L_{\rm data}$ to train PINN. Method-1 is abbreviated as M-1 and similarly the others.}
    \label{table_methods}
\end{table}

The $t$-grid, $x$-grid and $v$-grid given in table \ref{table_methods} refers to the equidistant grid points used by VOICE to solve equation \ref{Vlasov_eq} and \ref{Poisson_eq} for the parameters given in table \ref{sp}. These methods are devised keeping in mind the amount of data we need to store for the inference from simulation.
In M-1 method, training points are chosen randomly from full phase space $(x,v,t)$ which needs all the data to be stored. In M-2 method, we randomly choose and fix $x$-grid, $v$-grid and $t$-grid points for training, so only data at these points are required for storage. This drastically reduces the disk space required for inference. In M-3 method, $(x,v)$ points are chosen for randomly chosen timesteps, which requires to store the data for randomly chosen timesteps for each $x$ and $v$-grid points. Of course, once the training is done all the stored data can be deleted and only trained model is stored. M-2 and M-3 methods differ when it comes to using the potential data to estimate $L_{PDE}$. M-3 method necessitates the storage of all data points corresponding to the potential in the $x$-grid but M-2 method only requires the storage of selectively chosen $x$-grid points. For example, if we run a simulation, we will have to wait and store all the data to select random points from all the $(x,v,t)$ phase space while using M-1. But using M-2 requires saving the data at the chosen $x$-grid, $v$-grid points at the random time steps, which reduces the disk space required for the training data. For using M-3, while running the simulation, data can be stored at random time steps but for all $(x,v)$. However, for M-2 since the grid-points for $x$ are fixed, the potential values are needed for only few $x$ positions.
We compare these methods to obtain the most effective way of choosing training points while using the least amount of stored data points for accurate inference. 

\subsection{Comparison of DeepNN and PINN to infer the stored simulation data}
\label{compare_deepNN_PINN}
In this section, we conduct comparative analysis of PINN and DeepNN in terms of their suitability for data storage in simulations. A subset from $t_i=0$ till $t_f=10$ of the case-II data set is chosen. As concluded from the previous subsection, we use a NN with 15 layers and 40 nodes each and $swish$ activation function for both DeepNN and PINN. The training is done till 10000 epochs for both networks using different percentage of stored distribution function data (or $f$-data). M-1 method is used to select collocation points to train PINN, 5\% of collocation points (totalling 2.31x$10^6$ points) are randomly chosen for most of the PINN trainings.

\par
\begin{figure}[ht]
    \centering
    \subfloat[]{\includegraphics[width=7cm]{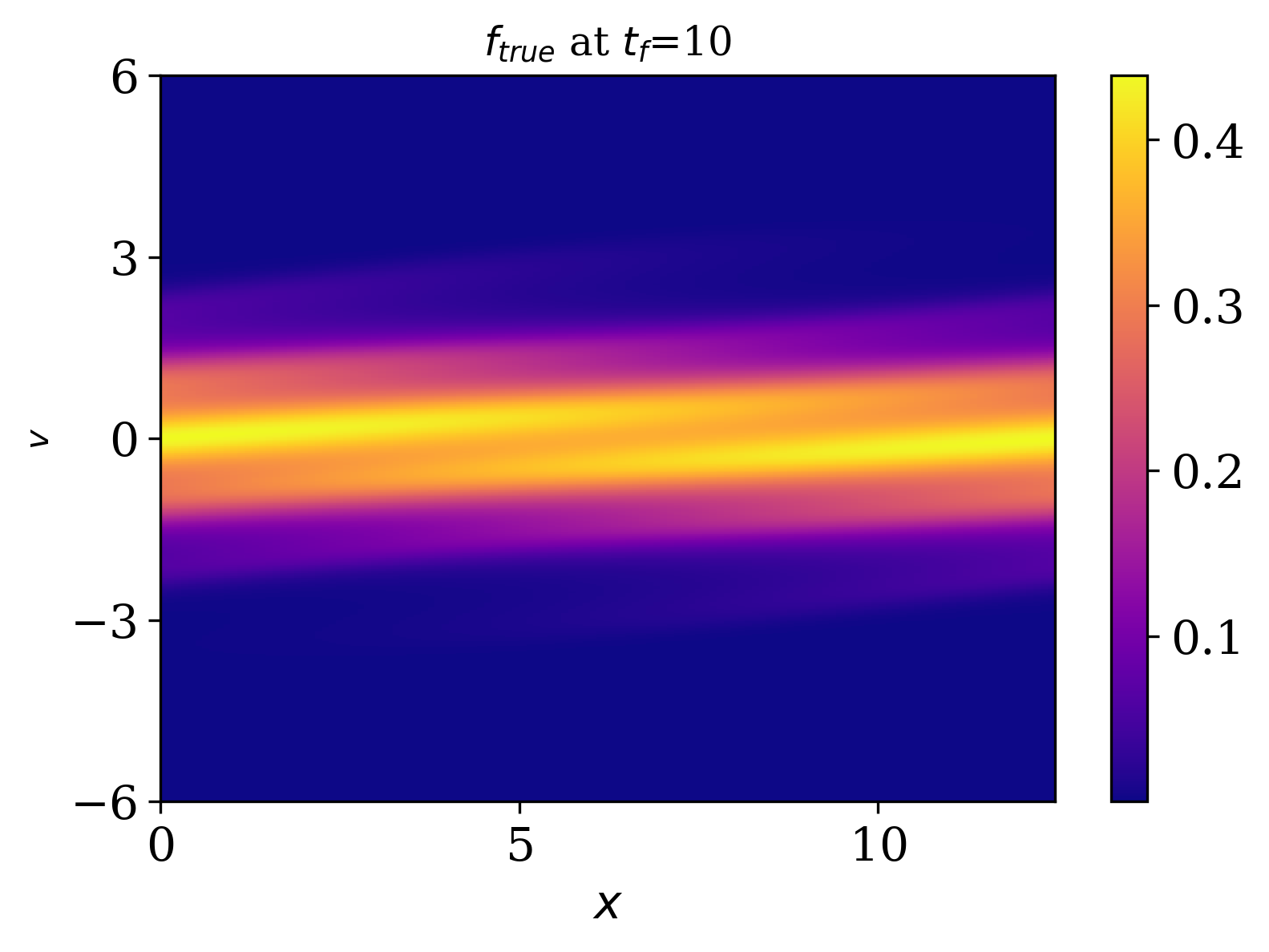}
    \label{true_error_2d_t10_landau}}
    \subfloat[]{\includegraphics[width=7cm]{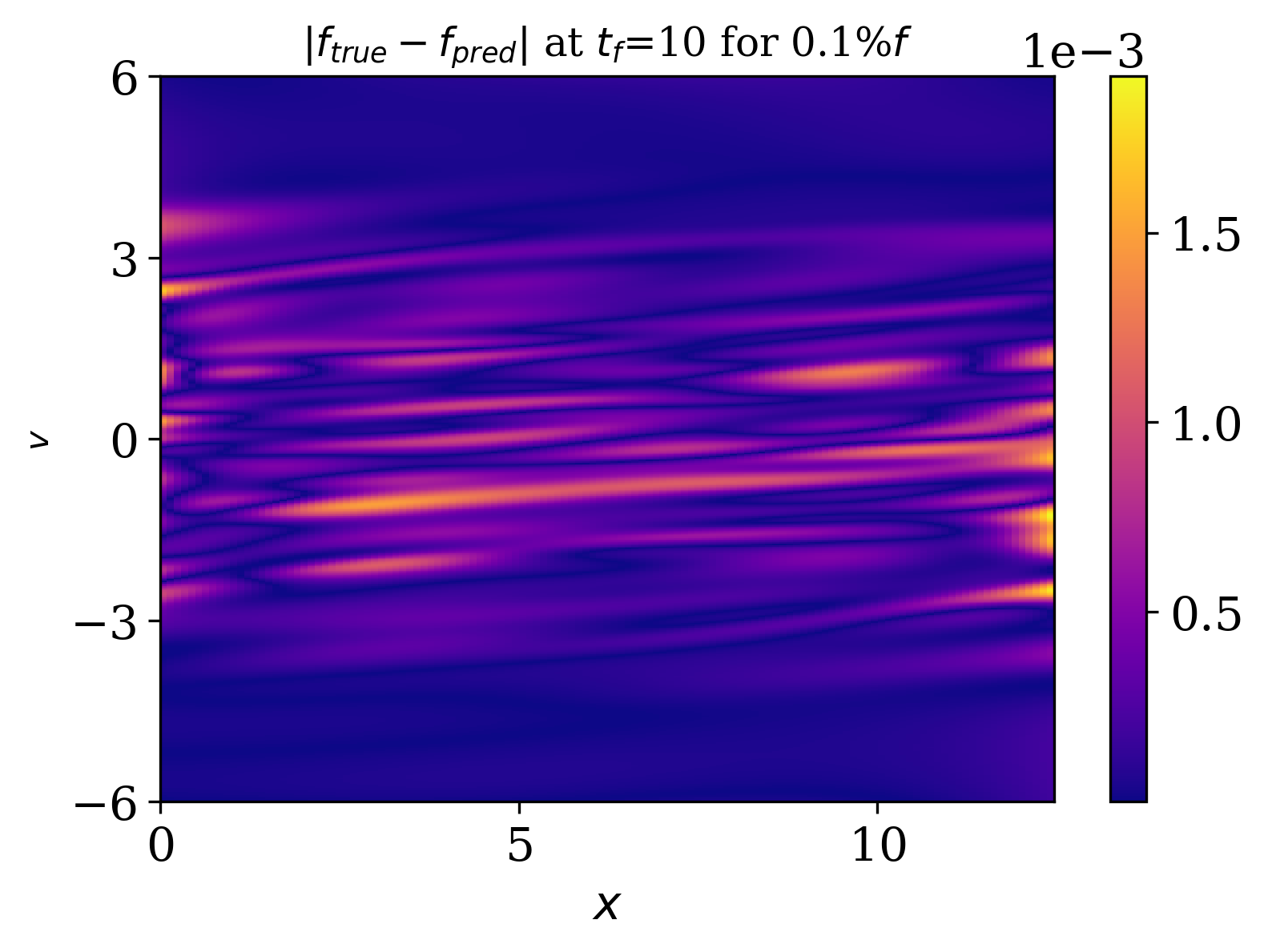}
    \label{error_2d_t10_landau_landau_NN_0p1f}}\\
    \subfloat[]{\includegraphics[width=7cm]{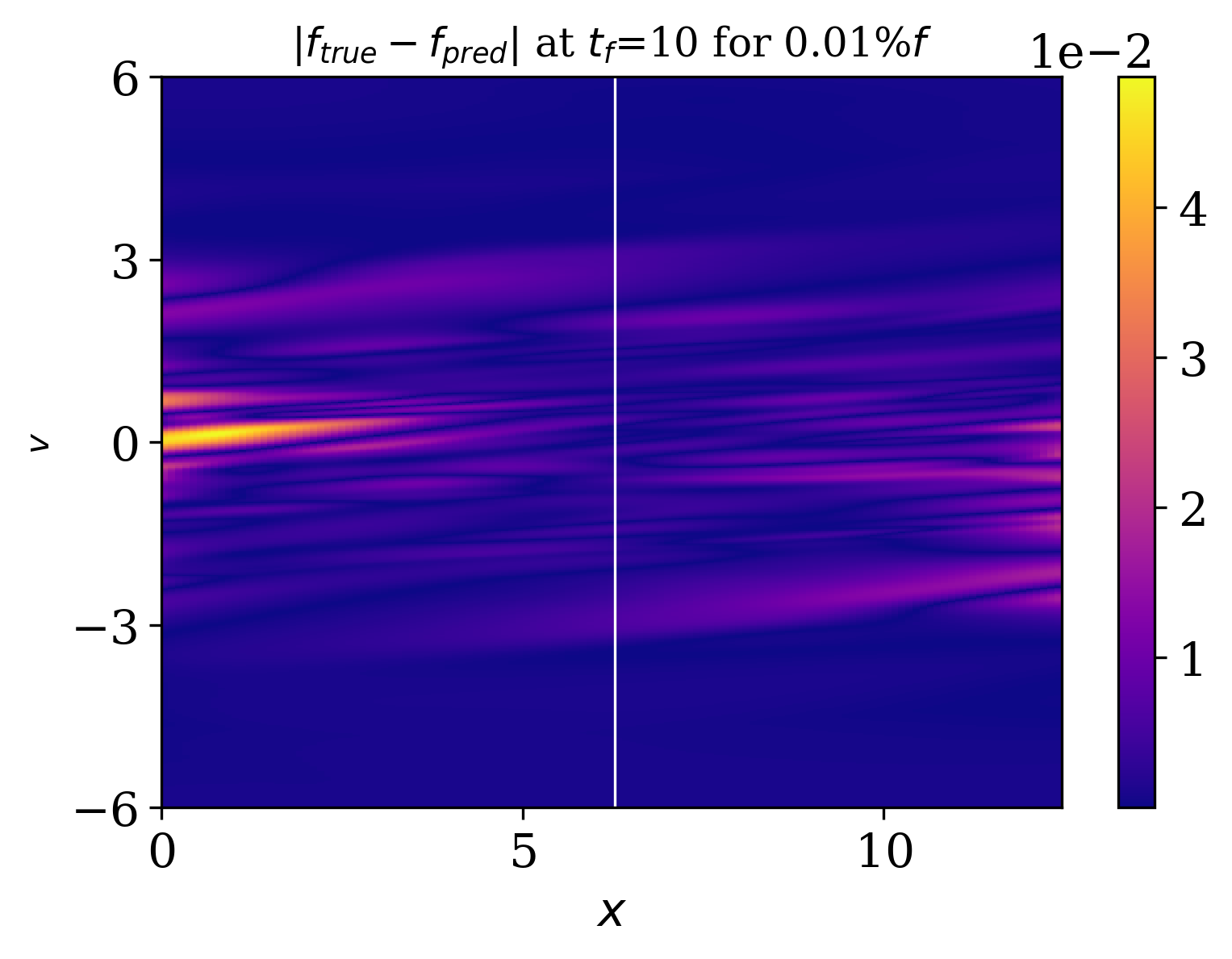}
    \label{error_2d_t10_landau_landau_NN_0p01f}}
    \subfloat[]{\includegraphics[width=7cm]{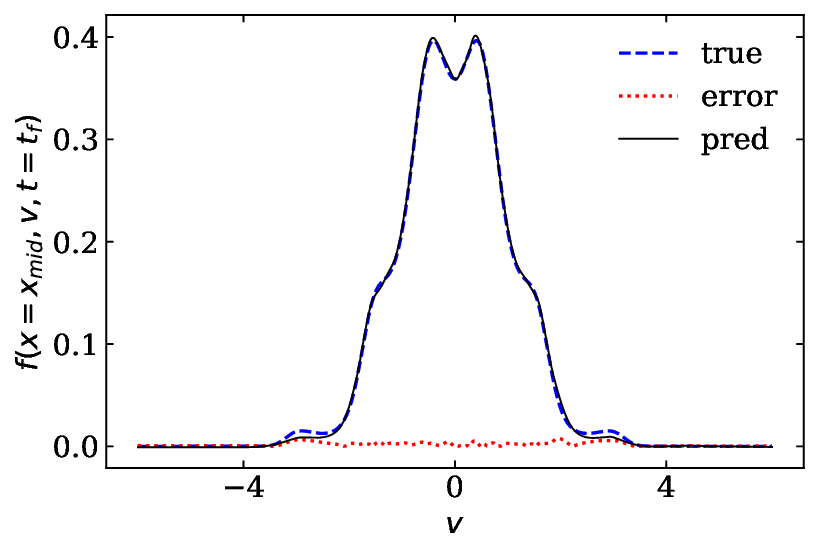}
    \label{error_1d_t10_xmid_landau_landau_NN_0p01f}}
    \caption{(a) The true 2D plot of distribution function, $f$ is plotted as a function of $x$ and $v$ at final time $t_f=10$ on the left. (b) The 2D plot of absolute error ($f_{true}-f_{pred}$) is plotted against $x$ and $v$ at the final time $t_f=10$ for the training done using DeepNN with 0.1\% of $f$-data as given in table \ref{table_loss_NN_model_percf}. (c) The absolute error for training done using 0.01\% $f$-data. (d) The true and prediction of $f$ are respectively plotted in blue dashed and black solid lines as a function of $v$ at final time and middle $x$-position, $x_{mid}$ (corresponding to the white solid line in (c)).}
    \label{model_comparision_nn_results}
\end{figure}

 The inference results of DeepNN and PINN trainings are given in the table \ref{table_loss_NN_model_percf} and \ref{randc1} for different percentage of chosen $f$-data. The Mean Square Error (MSE) given in the tables is the mean of square of the difference between true values of distribution function across all $\left(x,v,t\right)$ on which the training is done and the prediction by NN on the same points,
\beq
\text{MSE} = \sum_{\left(x,v,t\right)\in\Omega_{\rm data}}\frac{\left\vert f_{\rm data}- f_{\rm pred}\right\vert^2}{N_{\rm data}} 
\eeq
The final test loss is the final value of test loss after taking a moving average over 500 epochs. In the following, the absolute error is defined as
\beq
\text{Absolute Error} = \left\vert f_{data} - f_{pred}\right\vert
\eeq
where, $f_{data}$ represents the true value of $\df$ for a given $\left(x,v,t\right)$ and $f_{pred}$, the predicted values from NN at the same $\left(x,v,t\right)$ points. This absolute error is plotted in figure \ref{error_2d_t10_landau_landau_NN_0p1f} for the training where 0.1\% of $f$-data is retained and compared with the case where only 0.01\% of $f$-data is retained (figure \ref{error_2d_t10_landau_landau_NN_0p01f}). It can be seen that the absolute error increased one order from $10^{-2}$ to $10^{-3}$.
\begin{table}[ht]
    \centering
    \begin{tabular}{lllccl}
    \toprule
    {S.No.}&{\% $f$-data} & {Total training points}  &  {Final test loss} & {MSE} \\
    \midrule
    1& 1 & 461764   &  2.3x$10^{-07}$ &  7.34x$10^{-08}$\\
    2& 0.1 & 46176  &  3.30x$10^{-07}$ &  2.51x$10^{-08}$\\
    3& 0.05 & 23087 & 3.55x$10^{-06}$ &  3.31x$10^{-06}$\\
    4& 0.01 & 4617  &  5.25x$10^{-06}$ &  1.07x$10^{-05}$\\
    \bottomrule
    \end{tabular}
    \caption{Training results after 10000 epochs of training on 1 GPU using DeepNN model of 15 layers with 40 nodes each is given for different percentage of $f$-data used. Including the actual number of points used for training, the final test loss after taking the moving average over 500 epochs and MSE of prediction over the full dataset. Total time taken by each training is around 3.34 GPU hours.}
    \label{table_loss_NN_model_percf}
\end{table}

\begin{figure}[ht]
    \centering
    \subfloat[]{\includegraphics[width=7cm]{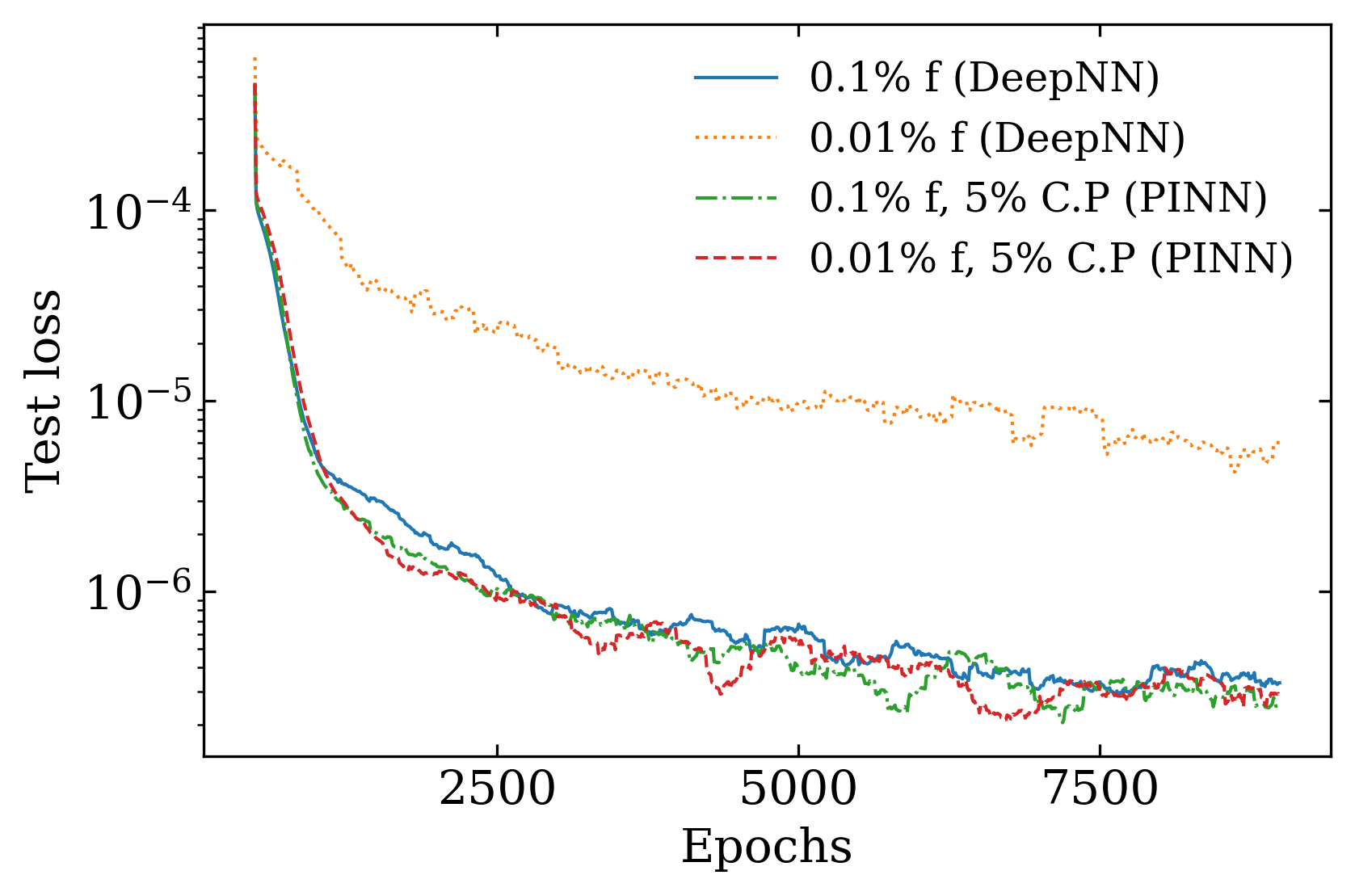}
    \label{PINN_NN_comparison_from_t0_till_t10}}
  \subfloat[]{\includegraphics[width=7cm]{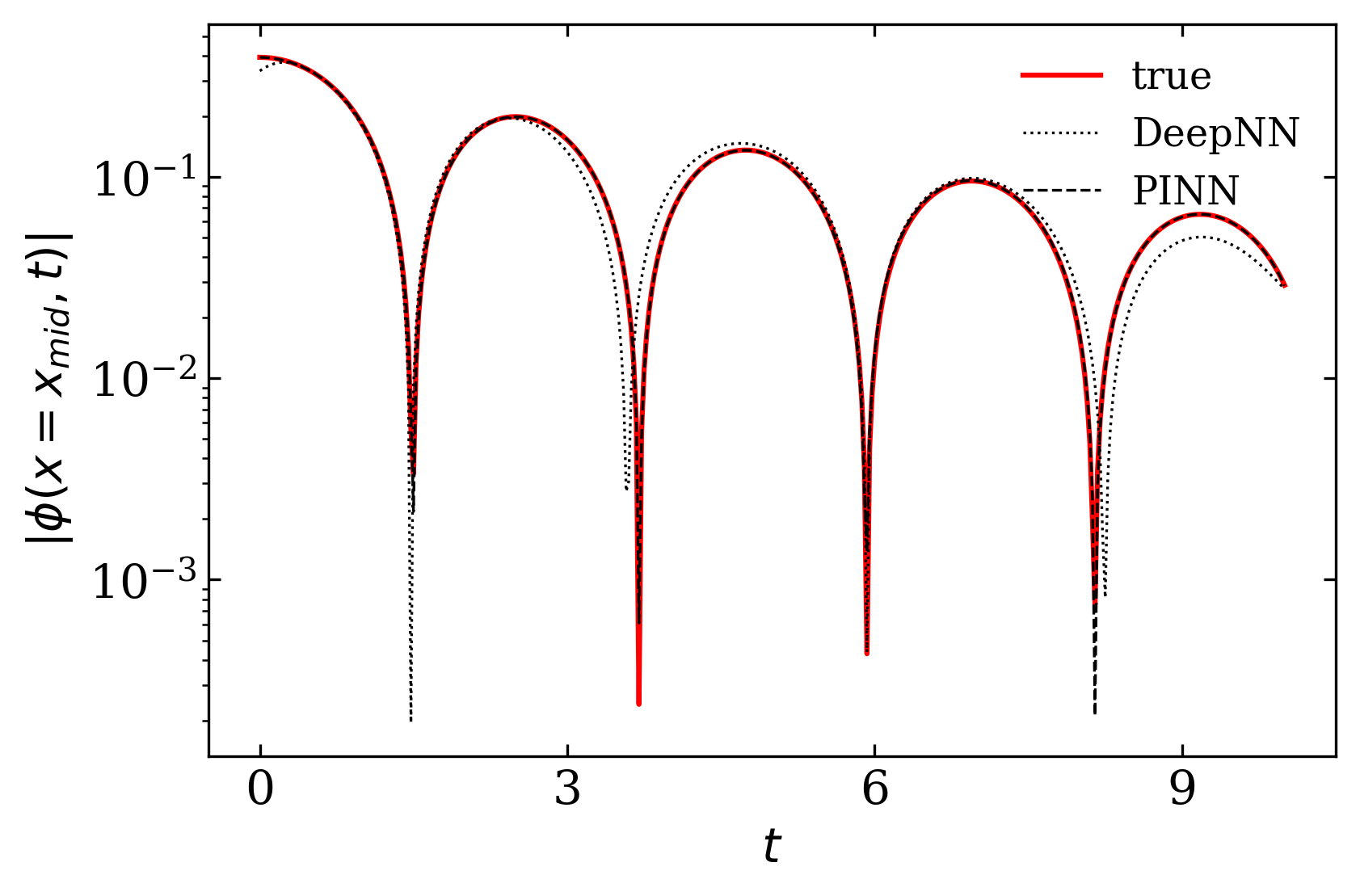}
    \label{PINN_NN_comparision_for_phi_till_t10_landau0p1}}
    \caption{(a) The training test loss is plotted as a function of number of epochs using the model with 40 nodes in each of the 15 layers on 1 GPU for PINN and DeepNN. The green dot dashed and red dashed lines are the test loss curve for PINN using 0.1\% and 0.01\% of $f$-data respectively,  also, utilizing 5\% collocation points (abbreviated as C.P) as per M-1 given in table \ref{table_methods}. The blue solid and orange dotted lines are the test loss curve for DeepNN using 0.1\% and 0.01\% of $f$-data. (b) The true electrostatic potential at the middle of $x$ is plotted in red solid line as a function of time. The prediction from PINN and DeepNN of $\phi$ for 0.01\% of $f$-data is plotted in black dashed and dotted lines respectively.}
    \label{comparision_pinn_nn}
\end{figure}

\begin{table}[ht]
    \centering
    \begin{tabular}[width=15cm]{llcll}
    \toprule
    {S.No.}&{\% $f$-data} & {\% Collocation points}  & {Final test loss} & {MSE} \\
    \midrule
    1& 1 &  5  & 3.01x$10^{-07}$ & 1.03x$10^{-07}$\\
    2& 1  & 10 & 2.71x$10^{-07}$ & 8.83x$10^{-07}$\\
    3& 0.1 & 5 & 2.21x$10^{-07}$ & 4.31x$10^{-08}$\\
    4& 0.05 & 5 & 1.94x$10^{-07}$ & 3.47x$10^{-08}$\\
    5& 0.01 & 5 & 2.20x$10^{-07}$ & 4.30x$10^{-08}$\\
    6& 0.01 & 1 & 2.26x$10^{-07}$ & 3.65x$10^{-07}$\\
    \bottomrule
    \end{tabular}
    \caption{Training results after $\approx$10000 epochs of training on 1 GPU using PINN with 40 nodes in each of 15 layers for randomly choosing training points as per M-1. The time taken for each training is around $\approx$11 GPU hours. The total number of 1\% $f$-data points are 461764 and 5\% collocation points in total amounts to 2.31x$10^6$. The final test loss is given after taking the moving average over 500 epochs. MSE of the prediction over the full dataset is given in the last column.}
    \label{randc1}
\end{table}
The results of PINN training are given in table \ref{randc1}. It can be seen from the table that choosing different percentage of collocation points does not affect much the final test loss for the same fixed percentage of $f$-data used.
The comparison of test loss during training between PINN and DeepNN is given in figure \ref{PINN_NN_comparison_from_t0_till_t10}. It can be clearly seen from the figure that the test loss for PINN training using 0.1\% $f$-data is the same as for DeepNN training but an order of magnitude smaller for 0.01\% $f$-data. This means that DeepNN is better suited than PINN for inference when 0.1\% $f$-data is used as it is less expensive to train. But PINN is clearly more powerful when less data (0.01\% $f$-data) is used. It also indicates that incorporating the information of the physical system in $L_{\rm PDE}$ helps infer the missing results even if we decrease the data used in training (from 0.1\% to 0.01\% of $f$-data).
In order to compare the difference in terms of physics between the training of two NNs, the electrostatic potential, $\phi$ at $x=x_{mid}$, is evaluated using the predicted distribution function values of PINN and DeepNN. A numerical technique using Fast Fourier Transform (FFT) is used to solve the Poisson equation as given in \ref{Poisson_eq}, details of which can be found in \cite{BOURNE2023112229}. The potential at $x=6.28$ (middle $x$-position) is plotted against time in figure \ref{PINN_NN_comparision_for_phi_till_t10_landau0p1}. It can be observed from the plots that the PINN prediction of $\phi$ overlaps with the true results. On the other hand DeepNN prediction falls short in case of 0.01\% $f$-data. It is to be noted that the determination of the total number of epochs for training is based on the accuracy of the predicted potential. It is observed that when the test loss reaches the order of 10$^{-7}$, the predicted values of $\phi$ align with the physical accuracy. Consequently, the number of epochs is adjusted accordingly during the subsequent training process.
These results confirm that PINN is able to capture the physical aspect of the simulated data and is more powerful than a simple DeepNN for inference when less data is used. More PINN training results using different selection methods as given in table \ref{table_methods} are discussed in \ref{a_1_infer}.
\begin{figure}[ht]
    \centering
    \subfloat[]{\includegraphics[width=7cm]{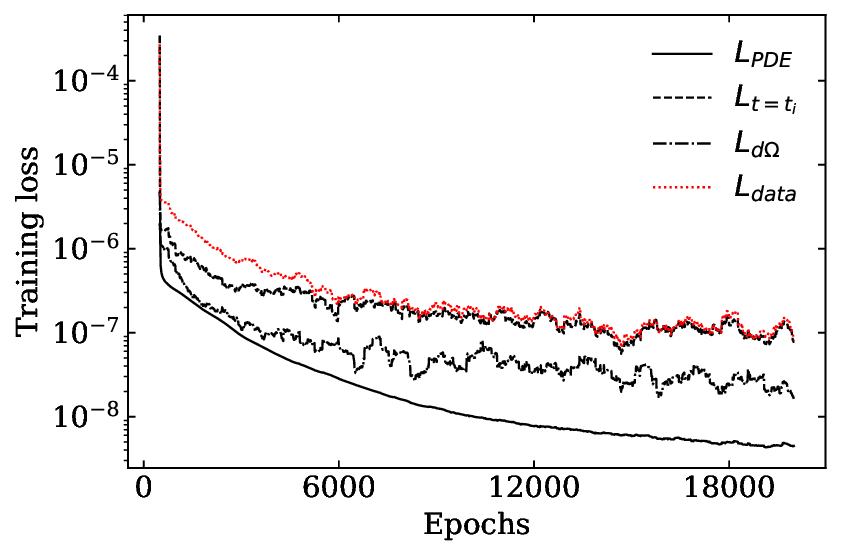}
    \label{landau_ep0p01_training_loss}}
    \subfloat[]{\includegraphics[width=7cm]{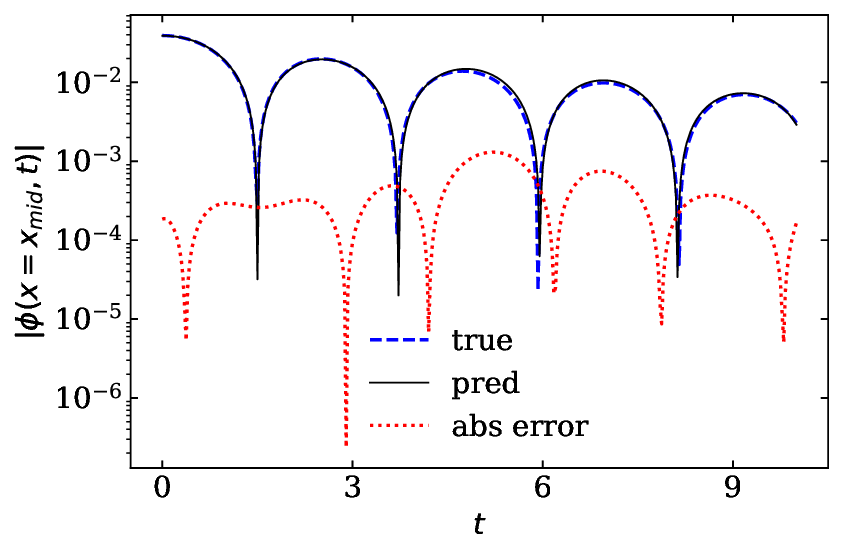}
    \label{landau_ep0p01_phi}}\\
    \subfloat[]{\includegraphics[width=7cm]{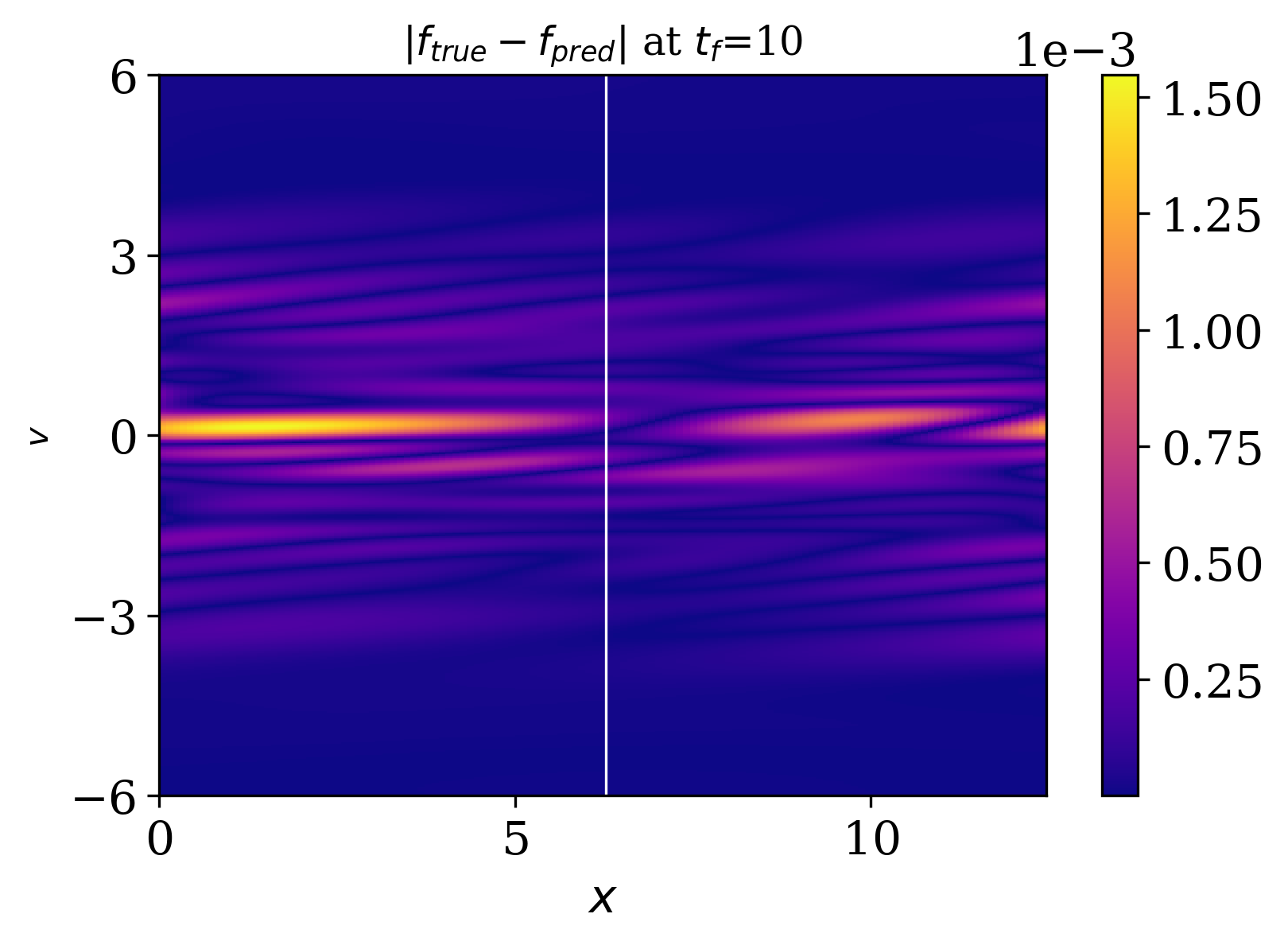}
    \label{landau_ep0p01_error_2d_tf}}
    \subfloat[]{\includegraphics[width=7cm]{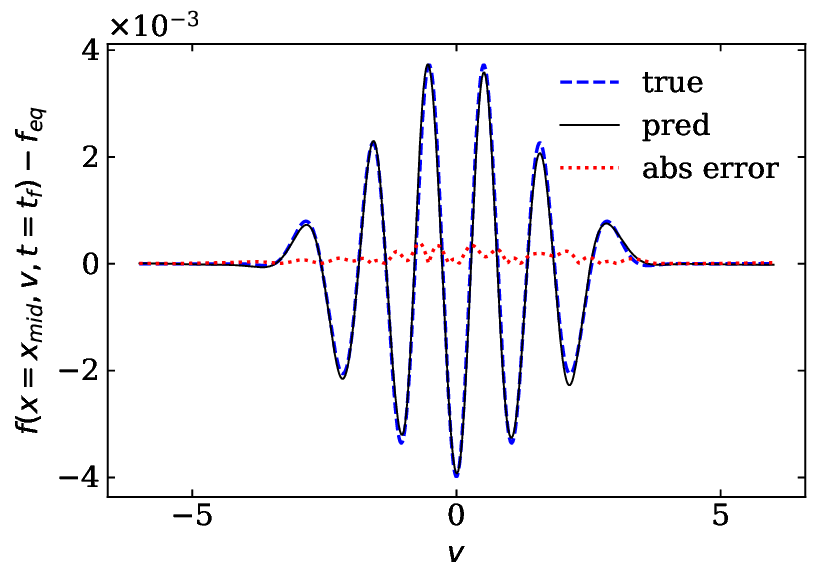}
    \label{landau_ep0p01_true_pred_1d_pert_tf}}
    \caption{Results of the training using PINN and the predictions compared to the true results for the Landau damping with $\epsilon=0.01$. (a) Evolution of the different terms of the loss function during the training. Each term has been averaged over 500 epochs in order to smooth the oscillations due to the gradient descent. The loss for the Vlasov equation, $L_{\rm PDE}$ is represented by solid black line. The loss for initial condition, $L_{t=t_i}$ for $t_i$=0 is given in black dashed line. The loss for all the boundary conditions, which includes, periodic and Neumann boundary conditions is shown in black dashdotted line. The loss for original data, $L_{\rm data}$ is shown in red dotted line. (b) For the the mid $x$-position, the electric potential at $t=t_f$ predicted by PINN (solid black line) is plotted together with the true value (dashed blue line), showing a good quantitative agreement. For more clarity, absolute error between the predicted and true values of $\phi$ is also plotted in red dotted line. (c) The absolute error between true and predicted value are plotted in $(x,v)$ at final time, $t_f$=10. (d) $f-f_{eq}$ ($f_{eq}$ representing the equilibrium $\df$) is compared with true and predicted values at $t_f$=10 and $x=x_{mid}$ (corresponding to the white solid line plotted in (c)).}
    
    \label{landau_ep0p01_infer_results}
\end{figure}
\par Another PINN with the same architecture is trained for Case-I in table \ref{sp} for a subset of data from $t_i$=0 till $t_f$=10. The training points are the same as given in the sixth training of table \ref{rndc2}, using method M-2. The results after training for 20000 epochs during 22 GPU hours are given in figure \ref{landau_ep0p01_infer_results}. The evolution of the different terms of the loss function during the training is shown in figure \ref{landau_ep0p01_training_loss}. It is observed that the neural network is indeed learning the distribution function. The prediction of electrostatic potential at the mid $x$-position is given in figure \ref{landau_ep0p01_phi}, showing good quantitative agreement with true results. The absolute error (also shown in the same figure) is of the order of $10^{-4}$ which confirms the accurate prediction. As shown in figure \ref{landau_ep0p01_error_2d_tf} with the plotting of the absolute error for distribution function against $(x,v)$, the maximum error is of the order of 10$^{-3}$ at final time, $t_f$=10. 
The perturbation in $\df$ can be captured by calculating the difference between the evolved distribution function, $f(x,v,t)$, and the equilibrium distribution function, $f_{eq}$ (given in equation \ref{eq_f0}), at any time, namely $\delta f(x,v,t) = f(x,v,t)-f_{eq}$. In the case of Landau damping, the distribution function exhibits minimal changes due to the small amplitude of the perturbation so the good indication of prediction accuracy is the comparison of $\df$. The predicted and true $\delta f$ at the mid $x$-position and final time is given in figure \ref{landau_ep0p01_true_pred_1d_pert_tf}, it reveals an excellent agreement with an absolute error of 10$^{-4}$.
\par The training time to infer the distribution function is increased by a factor of 2 from Case-II to Case-I. This is because in order to capture small changes in the distribution function an accuracy of 10$^{-8}$ is required.

%% file: Inference_to_solveVlasov/solveVlasov.tex
\subsection{PINN to solve Vlasov equation}
\label{PINN_solve_Vlasov}
Our aim is to store as few data as possible from the simulation. This can be controlled by reducing $N_{\rm data}$, i.e. the number of points to estimate $L_{\rm data}$. However, it is to be noted that when $N_{\rm data}\rightarrow 0$, this is equivalent to solving the Vlasov equation, since the only information we have left is the initial and boundary conditions for the $\df$. This means that we effectively solve the Vlasov equation using PINN. However, the electric potential values obtained from the solution of the Poisson equation are still utilized to calculate the corresponding electric field. This electric field is then used to evaluate the loss for the PDE, $L_{PDE}$. The same PINN architecture and parameters are used here as in the previous section. The total loss now reads
\begin{equation}
    L = L_{\rm PDE} + L_{t=t_{i}} + L_{\partial\Omega_{N}} +  L_{\partial\Omega_{P1}\times\partial\Omega_{P2}}
    \label{loss_valsov}
\end{equation}
We use the same methods for selecting collocation points that we described in the previous section. To make comparison easier with previous section results, the same Case-II of non-linear Landau damping from $t_i=0$ to $t_f=10$ is used. We also start with the same percentage of collocations points for PINN training in each of the methods described in table \ref{table_methods}. The test loss after training PINN for 20000 epochs for M-1 are given in table \ref{table_solveVlasov_1}. The number of epochs to get the same order of accuracy as in the previous section is increased from 10000 to 20000. This is because $L_{\rm data}$ is not used for training, which makes PINN harder to converge to the solution. This also means that it takes twice the time in the training.
\begin{table}[ht]
    \centering
    \begin{tabular}[width=15cm]{ccc}
    \toprule
    {S.No.} & {\% Collocation points}  & {Final test loss} \\
    \midrule
    1 & 5 & 1.80x$10^{-07}$ \\
    2 & 1 & 7.30x$10^{-08}$ \\
    3 & 0.5 & 1.56x$10^{-07}$ \\
    \bottomrule
    \end{tabular}
    \caption{Training results after 20000 epochs using PINN model of 15 layers with 40 nodes each is given for randomly chosen 5\% (totalling 2.31x$10^6$ points) and 1\% collocation points using M-1. The time taken for training is around 24 GPU hours. The final test loss is given after taking the moving average over 500 epochs. The MSE is given for all the phase space in which the training is done.}
    \label{table_solveVlasov_1}
\end{table}
\par Comparing the results of table \ref{table_solveVlasov_1} and \ref{randc1} reveals that without using any $f$-data, PINN also gives the accurate results. The ability of PINN is exploited to successfully solve the Vlasov equation. This means we just need to store the data for electrostatic potential ($\phi$-data) to infer the stored results, which represents 512 times less memory than the full distribution function data. Using DeepNN to infer $f$ using only $\phi$ is impossible, which confirms that PINN is efficient in storing and inferring or predicting the simulation results. All the $\phi$-data is used in training when M-1 method is used for choosing collocation points, M-2 and M-3 provides better alternatives. The results using M-2 and M-3 methods are discussed in \ref{a_2_solveVlasov}.
\begin{figure}[ht]
    \centering
    \subfloat[]{\includegraphics[width=7cm]{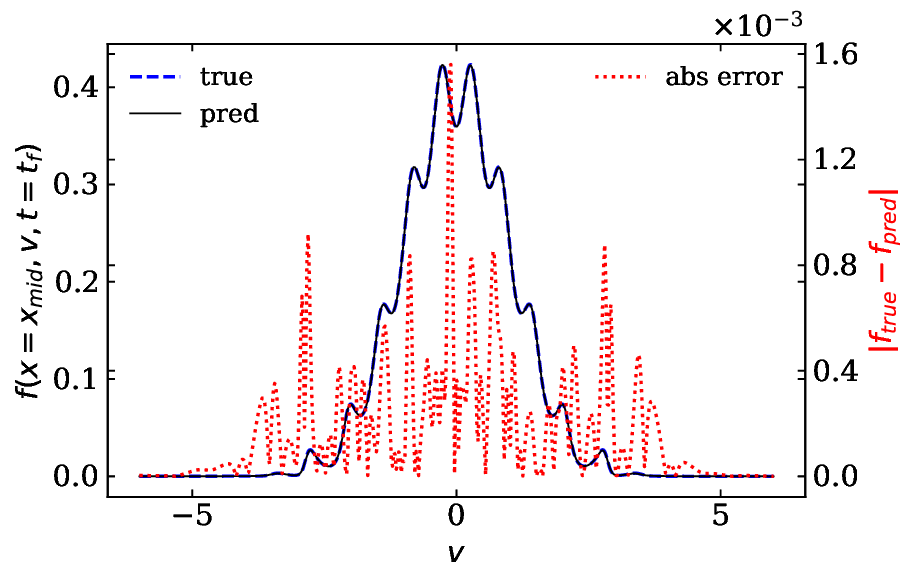}}
    \label{landau_solve_vlasov_t20_1d}
  \subfloat[]{\includegraphics[width=7cm]{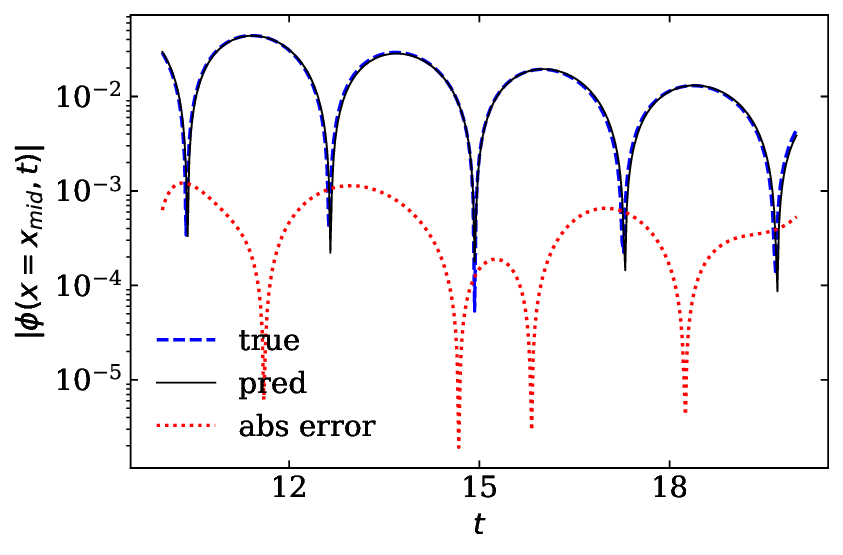}
    \label{bot_solve_vlasov_t20_phi}}
    \caption{(a) The prediction of distribution function is plotted in black solid line at final time, $t_f=20$ at mid $x$-position as a function of $v$ after training of 80000 epochs. The true value is plotted in blue dashed line and absolute error is plotted in red dotted line with y-axis on the right. (b) The true values of electrostatic potential at the mid of $x$-position are plotted in blue dashed line against time (y-axis is given in logarithmic scale). The $\phi$ evaluated from the prediction of distribution function by PINN is plotted in black solid line. The absolute error is plotted in red dotted line. The time taken for the training is 72 GPU hours.}
    \label{landau_solve_vlasov_t20}
\end{figure}

\par M-2 turns out to be the best method to choose collocation points. The lowest amount of data is used when 5\% of $t$-grid, 50\% of $x$-grid and 20\% of $v$-grid points are chosen for training to give the accuracy of 10$^{-7}$ (given in ninth training of table \ref{table_solveVlasov_2}). The total electric potential data points used from stored results are only 2560 (2.5\% of $\phi$), which amounts to 20 KiloBytes (KB) in disk space. The disk space required to store PINN is only 678 KB. In total, the disk space to store the simulation results from $t_i=0$ to $t_f=10$ of Case-II, is reduced from 400.8 Megabytes (MB) to 698 KB. The reduction in storage space required to store simulation data amounts to a significant factor of 588. This reduction signifies substantial decrease in the overall storage demands when employing PINN, thereby enabling more efficient data management and storage for the simulations.
\par From now on, the M-2 method is used for all the upcoming trainings. During the PINN training, we ended up using the $t$-grid points for every 20 timesteps, which makes the effective $dt$ = 0.25. Using the same PINN architecture, we solve the Vlasov equation for Case-II, from $t_i$=10 to $t_f$=20 for the sake of completeness. M-2 method is used to choose the collocation points: 5\% of uniformly chosen $t$-grid points,  50\% $x$-grid points and 50\% $v$-grid points. The results after 80000 epochs of training are given in figure \ref{landau_solve_vlasov_t20}. The predicted and true results exhibit excellent agreement. The increase in epochs from 20000 to 80000 is mainly because of the presence of highly non-linear data to be fitted, which takes more training time.
It is important to realise that we do not directly train PINN to solve Vlasov equation for the full time till $t_f=45$. This is because it is better to split the training for landau damping into a stretch of 10 unit times (from $t=0$ to $t=10$ then till $t=20$) per training since the non-linear phase takes a lot of time to converge for accurate prediction. One can predict correct results this way till $t_f=45$. This way we get accurate and faster results while keeping the data storage and resources consumed to minimum.
\par
\begin{figure}[ht]
    \centering
    \subfloat[]{\includegraphics[width=7cm]{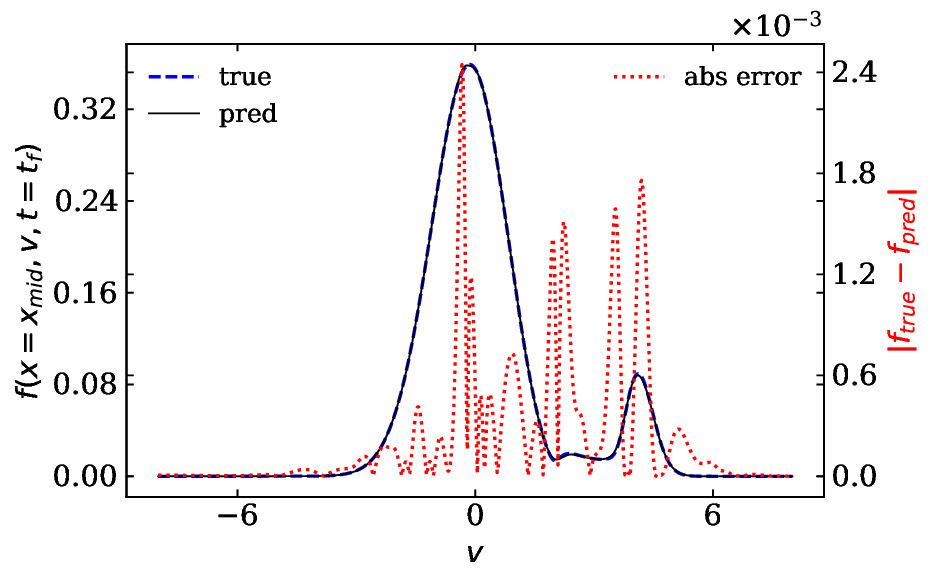}
    \label{bot_solve_vlasov_from_t55_t65_1d}}
  \subfloat[]{\includegraphics[width=7cm,height=4.5cm]{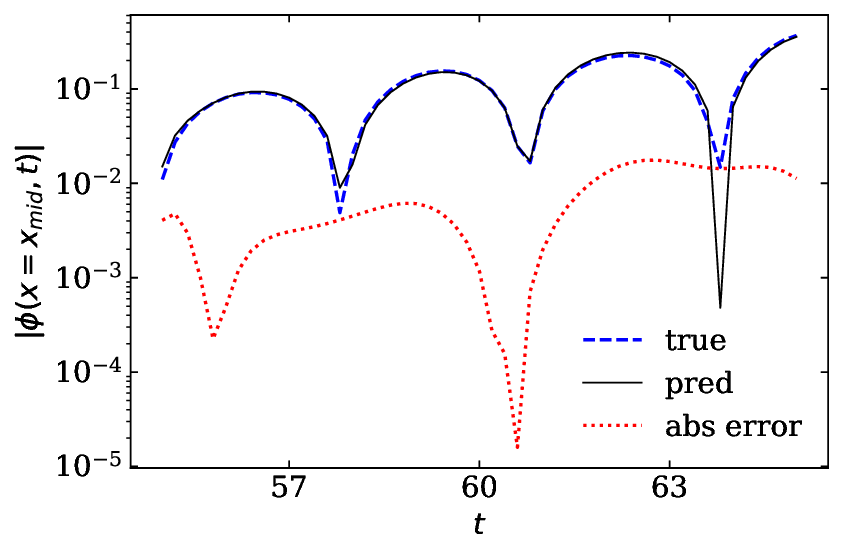}
    \label{bot_solve_vlasov_from_t55_t65_phi}}\\
    \subfloat[]{\includegraphics[width=14cm]{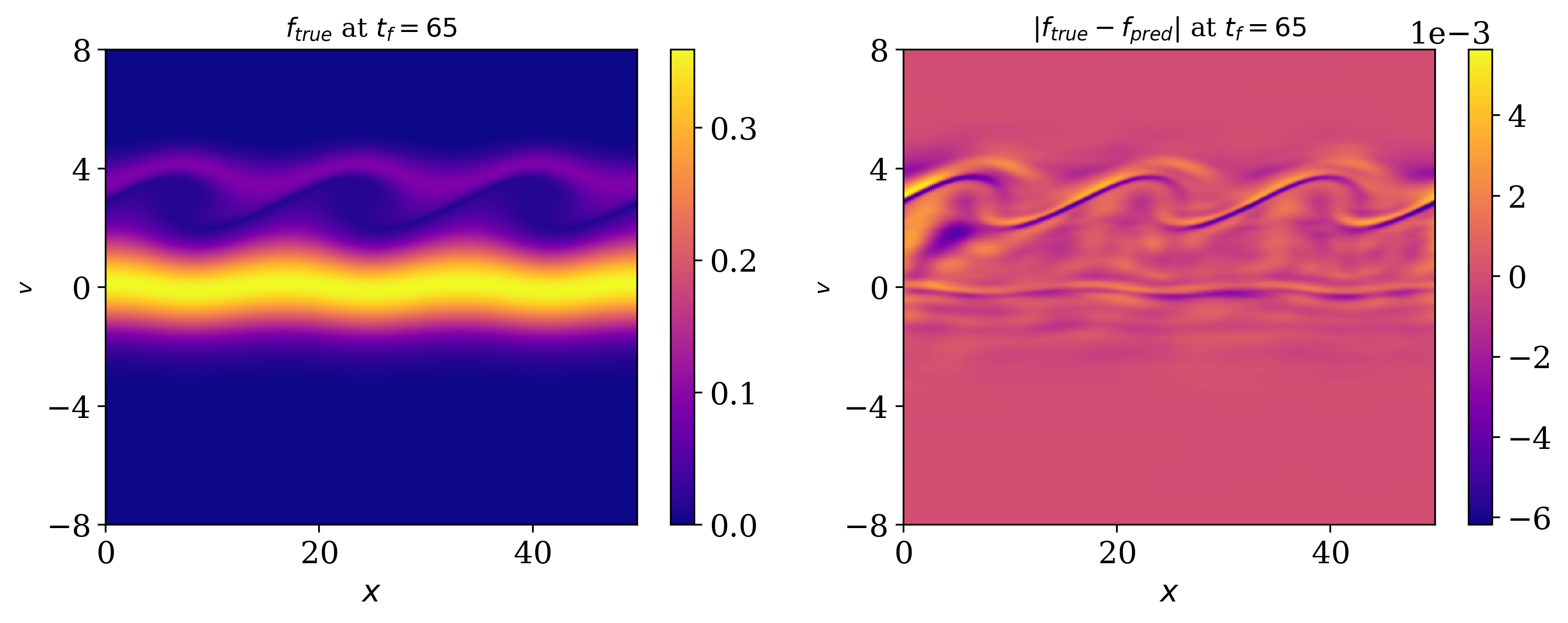}
    \label{bot_solve_vlasov_from_t55_t65_2d}}
    \caption{(a) The prediction of distribution function is plotted in black solid line at final time, $t_f=65$ for mid $x$ position as a function of $v$ after 70000 epoch training. The true value is plotted in blue dashed line and the absolute error is  plotted in red dotted line with y-axis in the right side. (b) The true electrostatic potential at the mid of $x$ position is plotted in blue dashed line as a function of time (y-axis is given in logarithmic scale). The prediction from PINN is plotted in black solid line and the absolute error is plotted in red dotted line. (c) The true distribution function is plotted over $x$ and $v$ on the left at final time $t_f$=65. The absolute error is plotted on the right for the same time against $x$ and $v$.}
    \label{bot_solve_vlasov_from_t55_t65}
\end{figure}

 We also solve the Vlasov equation for the bump-on-tail case as given in Case-III of table \ref{sp}. The subset of data used for training is from $t_{i} = 55$ till $t_{f} = 65$ because most activity happens in the simulation when it goes from linear to non-linear phase. The same M-2 method is used to choose collocation points for training, 50\% of $x$-grid points and 50\% of $v$-grid points for uniformly chosen 25\% of $t$-grid points. The $t$-grid points are chosen at every four time steps, which is equivalent to the effective $dt=0.2$. After 70000 epochs of training, the final test loss after taking the moving average over 500 epochs is 2.11x$10^{-7}$. The potential evaluated using the predicted distribution function values is in good agreement compared to the true results as seen from figure \ref{bot_solve_vlasov_from_t55_t65_phi}. It can be observed from figures \ref{bot_solve_vlasov_from_t55_t65_1d} and \ref{bot_solve_vlasov_from_t55_t65_2d} that PINN predictions are accurate to the order of $10^{-3}$.

%% file: SolvePoisson_to_IPINN/solvePoisson.tex
\section{Solving the Vlasov-Poisson system: implementation of an integrable PINN method to solve integro-differential equations.}
\label{fpinn_ipinn}
As explained before, since the Vlasov-Poisson system includes an integro-differential equation (IDE) one can not use PINN in its existing form to solve the whole system. There are some variants of PINN used to bypass this issue. Pang et al. showed in their paper \cite{f-PINNS} that one can solve IDEs combining PINN to solve the differential equation and a numerical method to solve the integral equation. They formulated an approximate function which depends on the PINN output. Such function includes trainable parameters and at the same time satisfies initial and boundary conditions. Using the approximate function, they solve the integral equation using the Finite Element (FE) method for the collocation points. The total loss is calculated by inserting the values of the solved integral equation into the loss for the PDE. PINN is afterwards is trained using this loss. This PINN variant is called fractional-PINN (f-PINN) as it is used to solve fractional differential equations. They used a particular setup of initial and boundary conditions for which an approximate solution can be easily formulated. Although the numerical technique they used is finite element method. They pointed out that f-PINN should work for any IDE with any initial and boundary conditions using any numerical method. 
In this section, f-PINN is used to solve the Vlasov-Poisson system and the limitation of the method is discussed. In addition, another method to solve IDEs using PINN, namely I-PINN (for Integrable-PINN), is formulated, which does not use any numerical method to solve the integral.
\subsection{f-PINN to solve Vlasov-Poisson equation}\label{fpinn_ide}
\begin{figure}[h!]
    \centering
    \subfloat[]{\includegraphics[width=7cm]{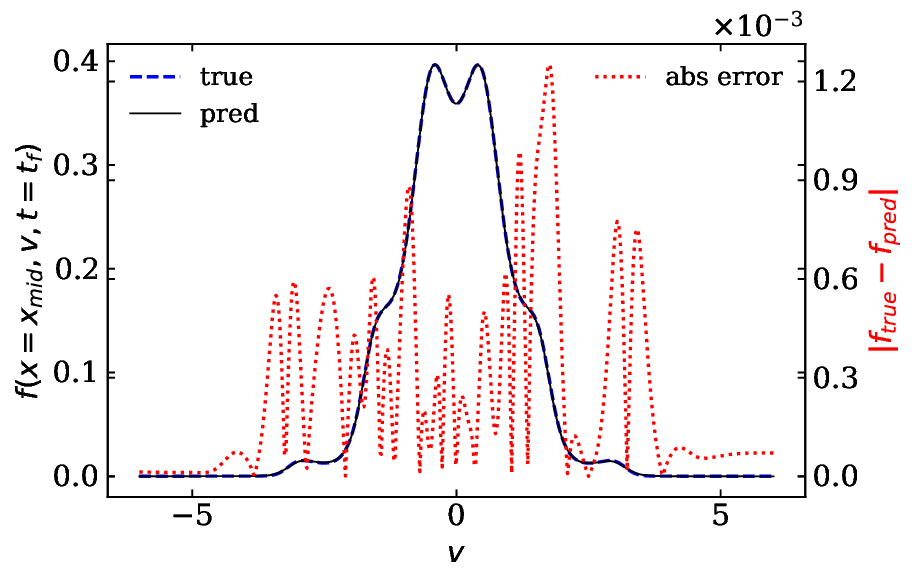}
    \label{landau_solve_poisson_till_t10_1d}}
    \subfloat[]{\includegraphics[width=7cm,height=4.5cm]{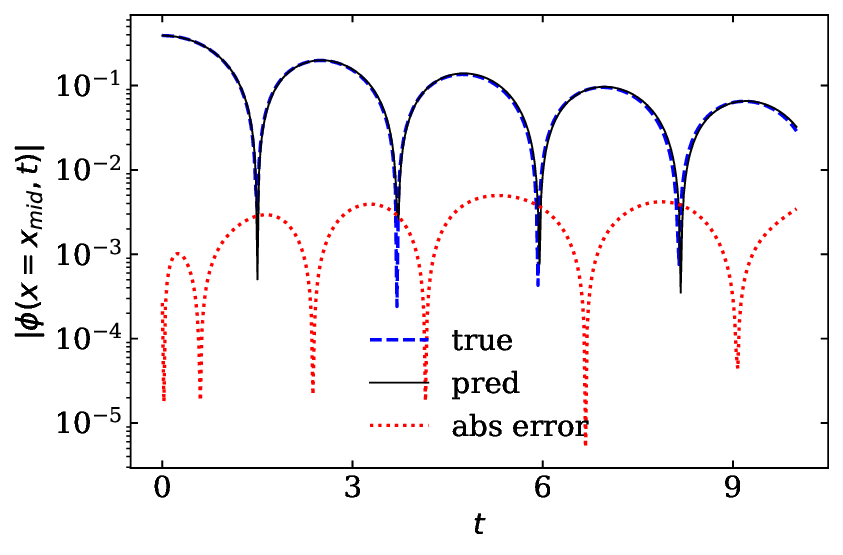}
    \label{landau_solve_poisson_till_t10_phi}}
    \caption{(a) The prediction of distribution function is plotted in black solid line at the final time, $t_f=10$, for the mid $x$-position as a function of $v$ after training of 30000 epochs using f-PINN. The true value is plotted in blue dashed line and the absolute error is plotted in red dotted line with the y-axis on the right. (b) The electrostatic potential at the mid $x$-position is plotted against time (y-axis is given in logarithmic scale). The true results from VOICE simulation are plotted in blue dashed line and the prediction from f-PINN is plotted in black solid line. The absolute error is plotted in red dotted line.}
    \label{landau_solve_poisson_till_t10}
\end{figure}

In contrast to the approach suggested by Pang et al., we do not rely on defining an approximate function based on initial and boundary conditions. Instead, we utilize loss terms to enforce the satisfaction of the initial and boundary conditions. Moreover, defining an approximate function becomes more challenging when working with periodic boundary conditions. Therefore, Fast Fourier Transform (FFT) is used instead as done in VOICE to solve the Poisson equation at each training instance. It is an additional part of training as compared to previous section and also the most resource consuming (even more than gradients) to evaluate during training. To solve the Poisson equation at the collocation points, equidistant points in $x$, $v$ and $t$ are defined. The grids resulted from uniform discretization are the same as those used in the VOICE simulation. The output from the network, $f_{\rm fPINN}\left(x,v,t;\boldsymbol{\theta}\right)$ is integrated over $v$ using the trapezoidal rule to obtain the gradient of the electric field.
\begin{equation}
\frac{\partial E\left(x,t;\boldsymbol{\theta}\right)}{\partial x}= 1 - \int_{v_i}^{v_f} f_{\rm fPINN} \left (x,v,t;\boldsymbol{\theta}\right)dv \quad \forall \left(x,t\right) \in \Omega_{\rm PDE}
\end{equation}
Using FFT one can write,
\begin{equation}
\texttt{fft}\left (E\left(x,t;\boldsymbol{\theta}\right)\right) = 
  \begin{cases}
    \frac{\texttt{fft}\left(\int_{v_i}^{v_f} f_{\rm fPINN} \left (x,v,t;\boldsymbol{\theta}\right)dv\right)}{\mathbf{\texttt{ik}}}, & \text{if } k\neq0 \\
    0, & \text{if } k = 0
  \end{cases}
\end{equation}
and taking Inverse Fast Fourier Transform (\texttt{ifft}) we obtain the electric field.
Here, $\mathbf{\texttt{i}}$ is the imaginary unit and $\texttt{k}$ is the wave vector. Note that now the electric field depends on trainable parameters ($\boldsymbol{\theta}$) of PINN. The loss for the PDE is changed from equation \ref{loss_valsov} as follows
\begin{align}
    &L_{\rm PDE}\left(\boldsymbol{\theta}\right)=\nonumber\\ 
    &\frac{1}{N_{\rm PDE}} \sum_{(x,v,t)\in\Omega_{{\rm PDE}}} \left\vert \frac{\partial f_{\rm PINN}\left(x,v,t;\boldsymbol{\theta}\right)}{\partial t}+ v \frac{\partial f_{\rm PINN}\left(x,v,t;\boldsymbol{\theta}\right)}{\partial x}+ E\left(x,t;\boldsymbol{\theta}\right)\frac{\partial f_{\rm PINN}\left(x,v,t;\boldsymbol{\theta}\right)}{\partial v} \right\vert^2 
\label{fpinn_loss}
\end{align}
\textit{Tensorflow-probability} library is used to evaluate $\texttt{fft}$, $\texttt{ifft}$ and the integral of $f$ using the trapezoidal rule.  During the training, in order to compute the gradient of the PDE loss with respect to $\boldsymbol{\theta}$, one has to compute the gradient of the electric field. This is different from what was done in the previous section, where the electric field is constant with respect to $\boldsymbol{\theta}$. This enables PINN to learn the relation between $f(x,v,t)$ and $E(x,t)$, making it f-PINN. The same subset from $t_i=0$ till $t_f=10$ of non-linear Landau damping simulation given in case-II is used for training. We found in previous section that method M-2 to select collocation points with (5\%,50\%,50\%) of $(x,v,t)$ is enough to get accurate results. We are using the same number of points to train f-PINN. The training is done till f-PINN has an accuracy of the order of 10$^{-7}$, which occurs after 30000 epochs. The time taken for training is 220 GPU hours due to the increased number of operations which includes solving the Poisson equation. The point to note here is that we are not using any pre-trained network or weighted loss or special optimizer which would help in speed up the training. Since our goal here is to test the feasibility of using PINN and its variants for plasma simulations, increasing the efficiency of training is not in the scope of the current work. The predictions from f-PINN training can be seen in figure \ref{landau_solve_poisson_till_t10}. The predicted and true distribution functions at final time and mid $x$-position are plotted as a function of $v$ in figure \ref{landau_solve_poisson_till_t10_1d}. It can be observed from the error plot in the figure that the value of absolute error is of the order of $10^{-3}$. The electric potential is evaluated from predicted distribution function, and the true and predicted values are plotted in figure \ref{landau_solve_poisson_till_t10_phi}, which completely overlap each other at mid $x$-position and the error is also of the order of $10^{-3}$.

After the training, the entire solution of the Vlasov-Poisson system under the form of weights and biases of the neural network is obtained without using any data from simulation except the initial and boundary conditions. However, using f-PINN implies that there is a grid dependence to solve the Poisson equation.

%% file: SolvePoisson_to_IPINN/IPINN.tex
\subsection{Using I-PINN to solve Vlasov-Poisson equation}
\label{ipinn_ide}
In this section we aim at solving integro-differential equations by keeping the features of PINN to have mesh-free solution without using any numerical technique. There are some very recent attempts to solve integro-differential equations. For example, Yuan et al. \cite{A-PINN} proposed an A-PINN (for Auxiliary-PINN) and apply to the Volterra equation. In practive, A-PINN outputs the solution of integral equation along with the solution of the PDE. Therefore, the total loss includes the loss for the PDE, for the initial and boundary conditions and for the integral equation. It means that A-PINN learns the solution of integral equation through the loss function. Moreover, works where integro-differential equations are solved in the framework of the Landau damping have been reported recently, making use of the so-called gPINN \cite{YU2022114823} and gPINNp \cite{datadrivenlandau}, where the kinetic data for the initial time is used for PINN training and the total loss is defined using the moments of the distribution function, essentially solving fluid equations.
The method we propose in the present paper is inspired by very recent works \cite{lindell2021autoint, kumar2022integral} and construct our own I-PINN (Integrable-PINN) capable for solving IDEs based on the fundamental theorem of Calculus. More specifically, let us assume that we need to compute the integral $\int_a^bf\left(v\right)dv$, with $f$ approximated by a neural network $f_{\rm NN}$. Instead of numerically integrating $f_{\rm NN}$, we define the function 
\begin{equation}
F\left(v\right) = \int_a^vf\left(v'\right)dv
\end{equation}
such that $F'\left(v\right) = f\left(v\right)$. Therefore, we can approximate $F$ by a neural network and use the automatic differentiation to train it so that its derivative is the integrand.
In other words, I-PINN does not learn the solution of the integral equation as A-PINN, instead it is satisfied through construction. No data from simulation is used to train I-PINN and the structure of the total loss remains the same as given in equation \ref{loss_valsov}. 

To apply I-PINN to the Vlasov-Poisson system given in equations \ref{Vlasov_eq} and \ref{Poisson_eq}, we need to define a function such that its gradient gives the $\df$. Two integrals have to be done over $x$ and $v$ to solve the Poisson equation. Therefore, we introduce a function $\hat{E}$ that we call grad network, since it gives $f$ after taking partial derivatives with respect to $x$ and $v$, i.e.
\beqa
& &\partial_x\partial_v \hat{E}\left(x,v,t;\boldsymbol{\theta}\right) =  f\left(x,v,t;\boldsymbol{\theta}\right)\nonumber\\ 
&\text{with, } &\hat{E}\left(x,v,t;\boldsymbol{\theta}\right) = \int_{v_i}^{v} \int_{x_i}^{x} f\left(x',v',t;\boldsymbol{\theta}\right) dx' dv'\nonumber\\ 
\label{IPINNdefination}      
\eeqa
assuming that $\hat{E}$ vanishes at the boundary. This condition is also taken into account in the total loss by adding an extra term for $\hat{E}$, called $L_{\hat{E}}$. The advantage of changing the variable is that there is no more integral in Poisson equation. Indeed, integrating the Poisson equation w.r.t $x$ leads to
\begin{multline}
    E\left(x,t;\boldsymbol{\theta}\right) = E\left(x_i,t;\boldsymbol{\theta}\right) + (x - x_i) + \\ \left(\hat{E}\left(x,v_i,t;\boldsymbol{\theta}\right) - \hat{E}\left(x_i,v_i,t;\boldsymbol{\theta}\right) \right) - \left (\hat{E}\left(x,v_f,t;\boldsymbol{\theta}\right) - \hat{E}\left(x_i,v_f,t;\boldsymbol{\theta}\right)\right)
\end{multline}\label{ipinn_E}

The total loss can then be defined with an extra term for the boundaries of $\hat{E}$,
\begin{equation}
\centering
    L = L_{\rm PDE} + L_{t=t_{i}} + L_{\partial\Omega_{N}} +  L_{\partial\Omega_{P1}\times\partial\Omega_{P2}} + L_{\hat{E}}
\label{IPINNloss}
\end{equation}
where,
\begin{align}
L_{\rm PDE}\left(\boldsymbol{\theta}\right)&=\frac{1}{N_{\rm PDE}} \sum_{(x,v,t)\in\Omega_{{\rm PDE}}} \bigg | \frac{\partial }{\partial x}\frac{\partial }{\partial v}\frac{\partial }{\partial t} \hat{E}\left(x,v,t;\boldsymbol{\theta}\right)+ \nonumber\\ 
&v \frac{\partial }{\partial v}\frac{\partial^2 }{\partial x^2} \hat{E}\left(x,v,t;\boldsymbol{\theta}\right)+\nonumber\\
& E(x,t)\frac{\partial }{\partial x}\frac{\partial^2 }{\partial v^2} \hat{E}\left(x,v,t;\boldsymbol{\theta}\right) \bigg |^2 \nonumber\\
L_{t=t_{i}}\left(\boldsymbol{\theta}\right)&=\frac{1}{N_{0}}\sum_{(x,v,t)\in\Omega_{0}}\left\vert\frac{\partial}{\partial x}\frac{\partial }{\partial v}\hat{E}\left(x,v,t;\boldsymbol{\theta}\right)-f_{0}\left(x,v,t\right)\right\vert^2 \nonumber\\
L_{\partial\Omega_D}\left(\boldsymbol{\theta}\right)&=\frac{1}{N_{\rm \partial\Omega_{D}}} \sum_{(x,v,t)\in\partial\Omega_{D}} \left\vert \frac{\partial}{\partial x}\frac{\partial }{\partial v}\hat{E}\left(x,v,t;\boldsymbol{\theta}\right)-f_{\partial\Omega_D}\left(x,v,t;\boldsymbol{\theta}\right)\right\vert^2\nonumber\\
L_{\partial\Omega_N}\left(\boldsymbol{\theta}\right)&=\frac{1}{N_{\rm \partial\Omega_{N1}}} \sum_{(x,v_i,t)\in\partial\Omega_{N1}}\left\vert \frac{\partial }{\partial x}\frac{\partial^2 }{\partial v^2} \hat{E}\left(x,v,t;\boldsymbol{\theta}\right)\right\vert^2 + \nonumber \\
&\frac{1}{N_{\rm \partial\Omega_{N2}}} \sum_{(x,v_f,t)\in\partial\Omega_{N2}}\left\vert \frac{\partial }{\partial x}\frac{\partial^2 }{\partial v^2} \hat{E}\left(x,v,t;\boldsymbol{\theta}\right)\right\vert^2 \nonumber\\
L_{\partial\Omega_{P1}\times\partial\Omega_{P2}}\left(\boldsymbol{\theta}\right)&=\frac{1}{N_{\rm \partial\Omega_{P}}}\sum_{\left((x_i,v,t),(x_f,v,t)\right)\in\partial\Omega_{P1}\times\partial\Omega_{P2}}\left\vert \frac{\partial }{\partial x}\frac{\partial }{\partial v}\hat{E} \left(x_i,v,t;\boldsymbol{\theta}\right)- \frac{\partial }{\partial x}\frac{\partial }{\partial v}\hat{E}\left(x_f,v,t;\boldsymbol{\theta}\right)\right\vert^2 \nonumber\\
L_{\hat{E}}\left(\boldsymbol{\theta}\right)& = \frac{1}{N_t}\sum_{t\in N_{t}} \left \vert\hat{E}(x_i,v_i,t)\right\vert^2 + \frac{1}{N_{\Omega_{P1}}}\sum_{(x_i,v,t)\in \partial \Omega_{P1}}\left\vert\frac{\partial \hat{E}(x_i,v,t)}{\partial v}\right\vert^2 + \nonumber\\ 
&\frac{1}{N_{\Omega_{P1}}}\sum_{i=0}^{N_{\Omega_{N1}}}\left\vert\frac{\partial \hat{E}(x,v_i,t)}{\partial x}\right\vert^2 \nonumber
\end{align}
Here $N_t$ is the total time points. I-PINN is constructed with 40 neurons in each of 15 layers, having as input vector $(x,v,t)$, but with output $\hat{E}(x,v,t)$. Since there is no integration in any equation, we solved the integral equation with the help of grad network created. In that sense, we call it automatic-integration, as the integral is inherently computed by construction.

\begin{figure}[h!]
    \centering
    \subfloat[]{\includegraphics[width=7cm]{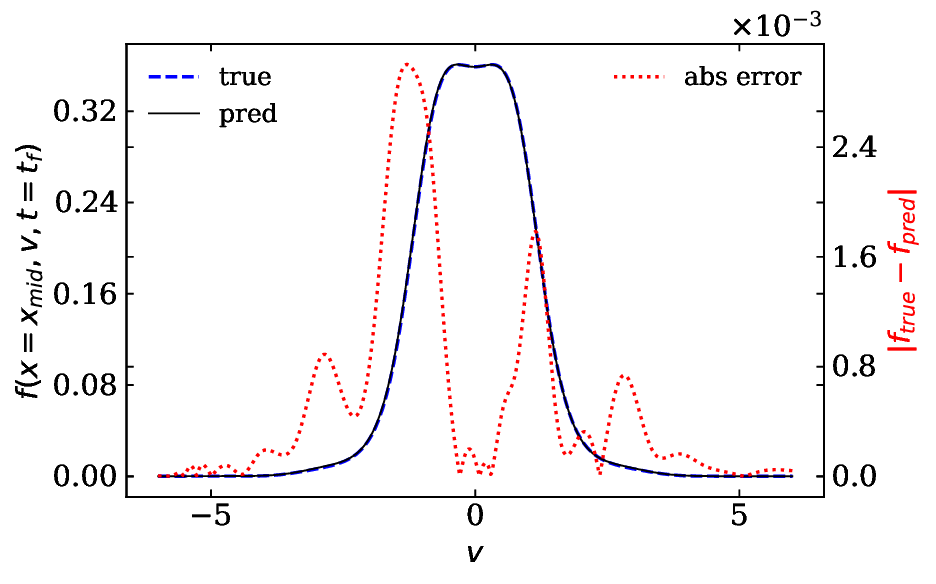}
    \label{landau_ipinn_till_t5_1d}}
  \subfloat[]{\includegraphics[width=7cm,height=4.5cm]{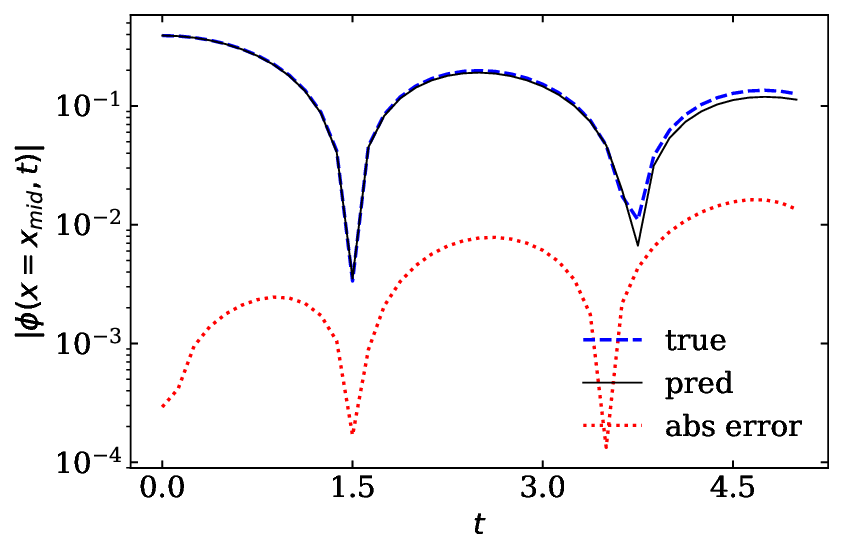}
    \label{landau_ipinn_till_t5_phi}}
    \caption{(a) The prediction of distribution function is plotted in black solid line  at final time, $t_f=5$ at mid $x$ position against $v$ after training of 100000 epochs using I-PINN model. The true value is plotted in red dashed line and the absolute error is plotted in red dotted line with values on the right side of y-axis. (b) The electrostatic potential ($\phi$) at the mid $x$ position is plotted against time (y-axis is given in logarithmic scale). The true results from VOICE simulation are plotted in blue dashed line and the prediction from I-PINN is plotted in black solid line. The absolute error is plotted in red dotted line.}
    \label{landau_ipinn_till_t5}
\end{figure}

I-PINN is now used to solve for the case-II of Landau damping as done before. Since, three fold gradients need to be computed at each epoch, which consumes resources and time, we decide to use a subset of small time interval, $t_i$=0 till $t_f$=5 for training I-PINN. The collocation points for training are chosen using M-2 method. The percentage of points in $t$,$x$ and $v$ are 20\%, 50\% and 50\%, respectively, totalling 1.19x10$^6$ points. The training was done on 4 GPUs for 116 hours. It is important to highlight here again is that the training is done without optimization of training time and resources consumed as it is a test bed case. The result after training is given in figure \ref{landau_ipinn_till_t5}. The predicted distribution function at $t_f$=5 at mid $x$ position overlaps with the true results within an error of the order $10^{-3}$ as given in figure \ref{landau_ipinn_till_t5_1d}. The electric potential evaluated from the predicted $f$ is given in figure \ref{landau_ipinn_till_t5_phi} at mid $x$-position, which also agrees with the true results within an error of the order of $10^{-3}$.

%% file: Conclusion/conclusion.tex
\section{Conclusion}\label{conclusion_futurework}
In this article, we explored the application of Physcis Informed Neural network (PINN) to the Vlasov-Poisson system for two purposes. First, as a storage method for kinetic data, since storing the weights and biases of a neural network requires much less disk space than storing the raw data. Second, as a mesh-free technique for integration of differential and integro-differential equations (IDEs). 
Vanilla PINN as it is, uses automatic differentiation which is not suitable for integro-differential equations. For this purpose we used f-PINN (PINN to solve Vlasov equation and FFT to solve Poisson equation). By doing so, we needed to define the grid to integrate the distribution function. This implies that the advantage of using PINN to give mesh-free results and without using numerical technique is lost. This motivated the development of I-PINN, a method to solve coupled integro-differential equations. I-PINN uses a grad network, where the integration is satisfied through construction based on the fundamental theorem of Calculus.

This work is also a proof of concept for I-PINN. In the future, we will focus on optimizing I-PINN to solve different integro-differential equations with the help of different techniques existing in the literature, for example using a weighted loss function, using different optimizers or adding causality information. We will also continue our work to use PINN as storage solution and expand it for higher dimensions and multi-species plasma simulations.

%% file: Appendix/appendix.tex
\section{Results for different selection methods of training points for inference using PINN}\label{a_1_infer}
\begin{table}[ht]
    \centering
    \begin{tabular}[width=15cm]{p{0.8cm}p{2cm}p{1.8cm}p{2cm}p{2cm}p{2cm}}
    \toprule
    {S.No.}&{$f$-data (\%t,\%x,\%v)} & {Total training points for $L_{data}$}  & {Collocation points (\%t \%x \%v)} & {Total training points for $L_{PDE}$} & {Final test loss} \\
    \midrule
    1 &(100,2,5) & 36045 & (100,20,25) & 2.31x$10^6$ & 3.08x$10^{-07}$ \\
    2 &(5,10,20) & 44064 & (5,100,100) & 2.36x$10^6$ & 2.82x$10^{-07}$ \\
    \midrule
    3 &(5,10,20) & 45165 & (5,100,100) & 2.42x$10^6$  & 2.17x$10^{-07}$ \\
    4 &(5,5,10) & 11291 & (5,50,100) & 1.21x$10^6$  & 2.70x$10^{-07}$ \\
    5 &(5,5,10) & 11291 & (5,50,50) & 6.05x$10^5$  & 2.05x$10^{-07}$ \\
    6 &(5,4,5) & 5120 & (5,40,50) & 4.81x$10^5$  & 3.32x$10^{-07}$ \\
    \bottomrule
    \end{tabular}
    \caption{Training results after 10000 epochs of training using PINN model of 15 nodes with 40 nodes each is given for M-2 method of choosing the different percentage of training points. The first two trainings (S.No. 1,2) of the table uses all the $t$-grid, $x$-grid and $v$-grid to choose the given percentage of collocation points and $f$-data points in the respective spaces. The rest of the training in table (S.No.3,4,5,6) uses $t$-grid points after every 20 timesteps, which means 5\% of the total timesteps are used but not randomly. The time taken for each training is $\approx$12 GPU hours.}
    \label{rndc2}
\end{table}
\par The training for inference using PINN is done using other methods of choosing the collocation points and compared. PINN is trained using M-2 for the same total number of points, 1\% of $f$-data and 5\% of collocation points, as used in M-1. So, randomly chosen 2\% of $x$-grid points and 5\% of $v$-grid points for 100\% (all) the timestep (or $t$-grid), totalling 1\% of $f$-data. Similarly, randomly choosing 20\% of $x$-grid points and 25\% of $v$-grid points for 100\% of $t$-grid points, totalling 5\% of collocation points. To understand the dependency of collocation points on timesteps, another training  was done using randomly choose 5\% of $t$-grid points with 10\% and 20\% of $x$ and $v$ grid points totalling the same 1\% of $f$-data points. Similarly, 5\% of collocation points divided into $(5\%,100\%,100\%)$ of in $(t,x,v)$ grids. The moving average test losses after 10000 epochs are given in the first and second training of table \ref{rndc2}. It can be seen in the table that the test losses are very similar, which means that $t$-grid can be reduced to $5\%$. The same percentage is kept in the next training as in the second training, but the $t$-grid points are chosen uniformly at every 20 timestep from the simulation. The test loss in this case also turns out to be very similar to before, which can be seen at third training results in the table. This implies that the data can be stored for  every 20 timestep for inference instead of random timestep points and PINN can still give accurate results. The amount of $f$-data used in the PINN training in total to only 0.01\% and 1\% of collocation points and got equally accurate results as seen in the last training given in table \ref{rndc2}. In this training, only 2\% of electric potential data is used.
\begin{table}[ht]
    \centering
    \begin{tabular}[width=15cm]{p{0.8cm}p{2cm}p{1.8cm}p{2cm}p{2cm}p{1.5cm}p{2cm}}
    \toprule
    {S.No.}&{$f$-data (\%t,\%(x v))} & {Total training points for $L_{data}$}  & {Collocation points (\%t, \%(x v))} & {Total training points for, $L_{PDE}$} & {Final test loss} \\
    \midrule
    1 &(100,0.1) & 45657 & (100,5) & 2.31x$10^6$& 3.13x$10^{-07}$ \\
    \midrule
    2 &(5,2) & 47273 & (5,100) & 2.37x$10^6$& 2.18x$10^{-07}$ \\
    \bottomrule
    \end{tabular}
    \caption{Training results after 10000 epochs using PINN model of 15 nodes with 40 nodes each on 1 GPU is given for M-3 of choosing the training points. In the first training collocation points are chosen as, 100\% of $t$-grid points and randomly choosing 5\% of $(x,v)$ mesh-grid points for each of the $t$ point. 0.1\% of $f$-data $(x,v)$ points are randomly chosen for each $t$. The second training is done on $t$-grid points after every 20 timesteps, which means just 5\% of timesteps are used, uniformly. The $(x,v)$ percentage for collocation and $f$-data is chosen as 100\% and 2\% respectively. This is the reason that even if percentage of $(x,v)$ increases as compared to first, the number of points used in training remains almost the same. The time taken for training both the cases is around 12 GPU hours.} 
    \label{randc3}
\end{table}
M-3 method is applied by choosing 1\% of $f$-data points and 5\% collocation points for PINN training as done in M-2. It is done in two ways, choosing 0.1\% ($x,v$) points at each timestep for $L_{\rm data}$ and randomly choosing 5\% ($x,v$) points at each timestep for $L_{\rm PDE}$. Second, uniformly choose 5\% of $(x,v)$ grid points for $L_{\rm data}$ and all $(x,v)$ for $L_{\rm PDE}$ as collocation points for every 20 timestep. The results for both of the cases are given in table \ref{randc3} and have 10$^{-7}$ order of accuracy in test loss.

\section{Results for different selection methods of training points for solving Vlasov using PINN}\label{a_2_solveVlasov}
\begin{table}[ht]
    \centering
    \begin{tabular}[width=15cm]{p{1cm}p{2.5cm}p{2cm}p{2cm}p{2cm}}
    \toprule
    {S.No.}& {Collocation points (\%t \%x \%v)} & {Total training points for $L_{PDE}$}  & {Final test loss} \\
    \midrule
    1 & (100,20,25) & 2.31x$10^6$ & 2.98x$10^{-06}$ \\
    2 & (5,100,100) & 2.36x$10^6$ & 2.51x$10^{-07}$ \\
    \midrule
    3 & (5,100,100) & 2.42x$10^6$ & 2.83x$10^{-07}$ \\
    4 & (5,50,100) & 1.21x$10^6$ & 2.81x$10^{-07}$ \\
    5 & (5,50,50) & 6.05x$10^5$  & 2.85x$10^{-07}$ \\
    6 & (5,20,100) & 4.72x$10^5$  & 7.63x$10^{-05}$ \\
    7 & (5,100,20) & 4.81x$10^5$ & 2.79x$10^{-07}$ \\
    8 & (5,100,5) & 1.18x$10^5$  & 4.71x$10^{-05}$ \\
    9 & (5,50,20) & 2.40x$10^5$  & 2.59x$10^{-07}$ \\
    \bottomrule 
    \end{tabular}
    \caption{Training results after 20000 epochs of training using PINN model of 15 nodes with 40 nodes each is given for different percentage of collocation points using M-2. First two trainings (S.No. 1,2) of the table uses all the $t$-grid to choose the given percentage of points. The rest (S.No. 3,4,5,6,7,8,9) uses $t$-grid points after every 20 timesteps, which means actually using just 5\% of timesteps, uniformly. The training time decreases from ~24 to 20 GPU hours as the collocation training points are reduced. The final test loss is given after taking the moving average over 500 epochs.}
    \label{table_solveVlasov_2}
\end{table}
\begin{table}[ht]
    \centering
    \begin{tabular}[width=15cm]{p{1cm}p{2.5cm}p{2cm}p{2cm}p{2cm}}
    \toprule
    {S.No.}& {Collocation points (\%t, \%(x v))} & {Total training points for $L_{PDE}$} &  {Final test loss} \\
    \midrule
    1 & (100,5) & 2.31x$10^6$  &  1.92x$10^{-07}$ \\
    \midrule
    2 & (5,100) & 2.37x$10^6$  & 2.60x$10^{-07}$ \\
    \midrule
    3 & (5,50) & 1.1x$10^6$   & 2.24x$10^{-07}$ \\
    \midrule
    4 & (5,20) & 4.7x$x10^5$   & 2.69x$10^{-07}$ \\
    \bottomrule
    \end{tabular}
    \caption{Training results after 20000 epochs of training using PINN model of 15 nodes with 40 nodes each is given for different percentage of collocation points using M-3. The first training is done by choosing all the $t$-grid points and randomly choosing 5\% of $(x,v)$ points for each of $t$-grid point. The second training uniformly choose $t$-grid points after every 20 timesteps, which means using 5\% of total $t$-grid points.}
    \label{table_solveVlasov_3}
\end{table}
\par It can be seen from the comparison of test loss of first training with second training in the table \ref{table_solveVlasov_2} that in order to get accurate results more percentage of collocation points from $x$ and $v$ grids are needed. A fraction of points (5\%) in $t$-grid can be used which was also observed in previous section. The percentage of collocation points used are reduced from $x$ and $v$ grids keeping the points from $t$-grid, uniformly chosen to 5\%. The fifth and seventh training in the table reveals that at least 50\% of $x$-grid points and 20\% of $v$-grid points are needed to get accuracy of the order of 10$^{-7}$ in test loss. If lesser collocation points than this are used the accuracy of the order of 10$^{-5}$ is observed as given in sixth and eights training of table \ref{table_solveVlasov_2}, which is not accurate for physically relevant results. The ninth training uses the lowest number of points.
\par The collocation points used in the first trainings of M-2 and M-3 in respective tables is effectively 5\%. Although, after comparing the test losses, M-3 gives an order more accurate results. The reason is that PINN trains on different $(x,v)$ grid points at each time step in M-3 which gives more generality to training and makes it more accurate. The second training of M-3 uses 5\% of uniformly chosen $t$-grid points and all the $(x,v)$ points also gives equally accurate results. The amount of stored results for electrostatic potential used in first case of M-3 is 100\% but in second it is only 5\%. The difference between the M-2 and M-3 method can be observed here. The amount of $\phi$ used from stored results can be reduced to 2.5\% or less in M-2 case as done in table \ref{table_solveVlasov_2} because the percentage of $x$-grid points can be chosen separately. In M-1 however, 100\% points from $x$-grid is used as is not independently chosen.